\documentclass[11pt]{article}
\pdfoutput=1

\usepackage{amsmath,amssymb}
\usepackage{epsfig}
\usepackage[numbers,compress]{natbib}
\usepackage{hyperref}
\usepackage{color}
\usepackage{slashed}
\usepackage{placeins}

\usepackage{amsfonts}
\usepackage{graphicx, rotating}
\usepackage{epstopdf}
\usepackage{latexsym}
\usepackage{rotating}

\usepackage[font={small}]{caption}   

\def\cw{c_{\textrm w}}
\def\sw{s_{\textrm w}}

\def\Xtt{{X_{\hspace{-0.09em}\mbox{\scriptsize2}\hspace{-0.06em}{\raisebox{0.1em}{\tiny\slash}}\hspace{-0.06em}\mbox{\scriptsize3}}}}
\def\Xft{{X_{\hspace{-0.09em}\mbox{\scriptsize5}\hspace{-0.06em}{\raisebox{0.1em}{\tiny\slash}}\hspace{-0.06em}\mbox{\scriptsize3}}}}

\def\Xtt{{X_{\hspace{-0.09em}\mbox{\scriptsize2}\hspace{-0.06em}{\raisebox{0.1em}{\tiny\slash}}\hspace{-0.06em}\mbox{\scriptsize3}}}}
\def\Xft{{X_{\hspace{-0.09em}\mbox{\scriptsize5}\hspace{-0.06em}{\raisebox{0.1em}{\tiny\slash}}\hspace{-0.06em}\mbox{\scriptsize3}}}}
\def\Xet{{X_{\hspace{-0.09em}\mbox{\scriptsize8}\hspace{-0.06em}{\raisebox{0.1em}{\tiny\slash}}\hspace{-0.06em}\mbox{\scriptsize3}}}}
\def\Ytt{{Y_{\hspace{-0.09em}\mbox{\scriptsize2}\hspace{-0.06em}{\raisebox{0.1em}{\tiny\slash}}\hspace{-0.06em}\mbox{\scriptsize3}}}}
\def\Yft{{Y_{\hspace{-0.09em}\mbox{\scriptsize5}\hspace{-0.06em}{\raisebox{0.1em}{\tiny\slash}}\hspace{-0.06em}\mbox{\scriptsize3}}}}
\def\Yot{{Y_{\hspace{-0.09em}\mbox{\scriptsize-1}\hspace{-0.06em}{\raisebox{0.1em}{\tiny\slash}}\hspace{-0.06em}\mbox{\scriptsize3}}}}
\def\Ztt{{Z_{\hspace{-0.09em}\mbox{\scriptsize2}\hspace{-0.06em}{\raisebox{0.1em}{\tiny\slash}}\hspace{-0.06em}\mbox{\scriptsize3}}}}
\def\Zft{{Z_{\hspace{-0.09em}\mbox{\scriptsize-4}\hspace{-0.06em}{\raisebox{0.1em}{\tiny\slash}}\hspace{-0.06em}\mbox{\scriptsize3}}}}
\def\Zot{{Z_{\hspace{-0.09em}\mbox{\scriptsize-1}\hspace{-0.06em}{\raisebox{0.1em}{\tiny\slash}}\hspace{-0.06em}\mbox{\scriptsize3}}}}

\def\FLFR{{\bf{5}$_L$+\bf{5}$_R$\ }}
\def\FLOR{{\bf{14}$_L$+\bf{1}$_R$\ }}

\def\lra#1{\overset{\text{\scriptsize$\leftrightarrow$}}{#1}}

\definecolor{myred}{rgb}{0.7, 0, 0}
\definecolor{myblue}{rgb}{0, 0, 0.7}
\definecolor{mygreen}{rgb}{0.04, 0.7, 0.5}

\hypersetup{colorlinks,citecolor=myred,linkcolor=myblue,urlcolor=myblue,linktocpage=true}

 \def\be   {\begin{equation}}   \def\ee   {\end{equation}}
 \def\ba   {\begin{array}}      \def\ea   {\end{array}}
 \def\bea  {\begin{eqnarray}}   \def\eea  {\end{eqnarray}}
 \def\bean {\begin{eqnarray*}}  \def\eean {\end{eqnarray*}}
 \def\nn{\nonumber}
 \def\bry{\begin{array}}
 \def\ery{\end{array}}

\newcommand{\hhref}[1]{\href{http://arxiv.org/abs/#1}{arXiv:#1}}

\numberwithin{equation}{section}

\title{
\vspace{-2cm}
\begin{flushright}
\end{flushright}
\vspace{3cm}
\bf \LARGE
On Flavour and Naturalness of Composite Higgs Models
\vspace{.2cm}}

\date{}

\begin{document}

\author{Oleksii Matsedonskyi \\
\normalsize\itshape  Scuola Normale Superiore and INFN, Piazza dei Cavalieri 7, 56126 Pisa, Italy}

\maketitle

\begin{abstract}
We analyse the interplay of the constraints imposed on flavour-symmetric Composite Higgs models by Naturalness considerations and the constraints derived from Flavour Physics and Electroweak Precision Tests. 
Our analysis is based on the Effective Field Theory which describes the Higgs as a pseudo-Nambu-Goldstone boson and also includes the composite fermionic resonances. 
Within this approach one is able to identify the directions in the parameter space where the $U(3)$-symmetric flavour models can pass the current experimental constraints, without conflicting with the light Higgs mass. 
We also derive the general features of the $U(2)$-symmetric models required by the experimental bounds, in case of elementary and totally composite $t_R$. 
An effect in the $Z \bar b b$ coupling, which can potentially allow for sizable deviations in $Z \to \bar b b$ decay parameters without modifying flavour physics observables, is identified.        
We also present the analysis of the mixed scenario, where the top quark mass is generated due to Partial Compositeness while the light quark masses are Technicolor-like.  
\end{abstract}




\newpage

\tableofcontents

\newpage


\section{Introduction}

The discovery of the Higgs-like boson at 125~GeV by the LHC~\cite{beh} with no robust evidence for any type of deviations from the Standard Model (SM) predictions ruled out many models of New Physics, sharpened predictions of the others and at the same time stimulated the development of alternative approaches to the Hierarchy Problem. 
In particular, the discovery had important implications for the Composite Higgs models (CH)~\cite{gk,Contino:2003ve,Agashe:2004rs,Contino:2010rs}, in which the Higgs boson is screened from the UV physics by its composite nature, and separated from the other composite resonances due to the Nambu-Goldstone symmetry. The measured value of the Higgs mass in minimal CH models typically requires a presence of the colored composite fermionic resonances with a mass below 2~TeV~\cite{Contino:2006qr,Matsedonskyi:2012ym,Marzocca:2012zn,Pomarol:2012qf,Redi:2012ha,Panico:2012uw}. In these minimal scenarios, in contrast to the models where the Higgs potential is saturated by the SM-neutral states~\cite{twinhiggs}, one can expect for a large number of new physics signatures related to the direct production at the LHC~\cite{Contino:2008hi}, as well as to the indirect new physics probes, such as flavour and EWPT observables.


The main goal of this work is to consider the constraints imposed on Composite Higgs models with light composite colored fermions by flavour and electroweak precision tests (EWPT), and in particular to test the viability of different flavour patterns. Doing this, we want to concentrate on the implication of the pseudo-Nambu-Goldstone nature of the Higgs boson and the clash between the naturalness considerations, which often require at least a presence of light composite partners of the top quark, and the experimental constraints, pushing  the New Physics mass scale up. 

The strong dynamics of the underlying CH description is difficult to solve exactly. The first CH models, in their modern incarnation, were described in a dual 5-dimensional weakly coupled picture~\cite{Contino:2003ve,Agashe:2004rs}. Recently a set of purely four-dimensional UV completions for pseudo-Nambu-Goldstone boson (PNGB) Higgs was proposed~\cite{Caracciolo:2012je,Barnard:2013zea,Ferretti:2013kya}. 
An alternative approach to the problem is to distant from the concrete UV completions and to focus instead on the most necessary ingredients of the low-energy description~\cite{Contino:2006nn}.
Following this line in attempt to be as general as possible, but at the same time to account for the PNGB properties of the Higgs boson, we adopt the effective field theory approach driven by CCWZ rules~\cite{ccwz}. 
Using this approach, we will describe the most general implications of the global symmetry breaking in the strong sector, for which we choose the minimal one $SO(5)\to SO(4)$~\footnote{See~\cite{Gripaios:2009pe,Frigerio:2012uc,Mrazek:2011iu,Marzocca:2014msa} for less minimal choices allowing to generate two Higgs doublets or the Dark Matter candidate.}.
This approach was already widely used in the analyses of the top quark sector of the CH models (e.g.~\cite{Marzocca:2012zn,DeSimone:2012fs,ewpt}), and also of some particular problems of flavour physics (\cite{Azatov:2014lha,Antipin:2014mda}), while the more general flavour physics analyses were typically performed within the frameworks less sensitive to PNGB effects~\cite{Barbieri:2012tu,Redi:2012uj,KerenZur:2012fr}. 
Our discussion will be partially based on the previous study~\cite{Barbieri:2012tu}, being different in what concerns the implications of PNGB nature of the Higgs boson, choices of composite fermion multiplets and the mechanisms of quark mass generation.

The structure of the paper is the following. 
In Section~\ref{sec:frame} we introduce the general framework for our discussion -- CCWZ construction for CH model with one layer of composite fermions -- and describe different possible realizations of flavour.  
In Section~\ref{sec:pheno} we discuss in details the most important constraints on our scenarios derived from flavour physics and electroweak precision tests and identify the configurations of the parameter space minimizing deviations from the experimental measurements. 
Results of a combined numerical analysis are collected in Sections~\ref{num3} and~\ref{num2}, and Section~\ref{sec_sum} summarizes the results of this work.

\section{Framework}
\label{sec:frame}

The CH models under consideration can be seen as consisting of two sectors. The first one is the composite sector, containing the composite Higgs boson and other composite resonances. The Higgs is realized as a Nambu-Goldstone boson of the global symmetry $\cal G$ spontaneously broken by the strong sector condensate characterised by a scale $f_{\pi}$. In this work we take ${\cal G}=SO(5)$ broken to $SO(4)$, thus generating four goldstone bosons forming an $SU(2)_L$ Higgs doublet.
Hence the Higgs is exactly massless unless the strong sector is coupled to some source of an explicit $\cal G$-breaking.
The second sector of the model contains the elementary copies of all SM states except for the Higgs (and optionally for the right-handed top quark) transforming under the SM gauge symmetry group ${\cal G}_{\text{SM}} \subset {\cal G}$.
This elementary sector does not respect the full global symmetry $\cal G$, so once the two sectors are coupled, the one-loop effective potential generated by the elementary-composite interactions allows the Higgs to have a mass and fixes its vacuum expectation value (VEV) in a ${\cal G}_{\text{SM}}$-breaking direction.
The Lagrangian hence can be divided in the following parts
\be
{\cal L} = {\cal L}_{comp} + {\cal L}_{elem} + {\cal L}_{mix}\,.
\ee
The most appealing way to break the Goldstone symmetry, generate the Higgs mass and the top quark mass without introducing too large flavour-violating effects is provided by the mechanism of partial compositeness.
This mechanism postulates that the UV Lagrangian above the $\cal G$ symmetry breaking scale contains linear couplings between elementary fermions $q$ and  strong sector operators
\be
{\cal L}_{mix}^{\text{UV}} = \sum_q y \bar q {\cal O}_q \, ,
\label{eq:pc}
\ee
where the operators ${\cal O}_q$ transform in one of the $SO(5)$ representations. We will consider two representation choices, fundamental {\bf 5} and symmetric traceless {\bf 14}. In the first case both chiralities of the fermion $q$ will be assumed to have an elementary representative coupled to the strong sector. In the second case we will take the right-handed $q$ quark as a totally composite state, thus arising itself from the operator ${\cal Q}_q$ at low energies, and coupled to $q_L$ by means of the mixing~(\ref{eq:pc}). The two scenarios, with the  elementary quarks coupled to {\bf 5} and {\bf 14} will be called \FLFR and \FLOR respectively.

Our analysis will be based on the effective field theory (EFT) containing only the up-type quarks and their partners because they belong to the sector with the largest elementary-composite mixings, needed to generate the top mass. The large mixings imply that, on the one hand, at least some of the partners belonging to this sector, being involved in the Higgs mass generation, can not be too heavy~\cite{Contino:2006qr,Matsedonskyi:2012ym,Marzocca:2012zn,Pomarol:2012qf,Redi:2012ha,Panico:2012uw}. On the other hand they naturally introduce the largest deformations of the observables with respect to the Standard Model.  
Therefore one can expect that the phenomenological analysis based mostly on the up-type partners will reveal the dominant effects which are enforced by naturalness. It will allow to squeeze the parameter space of the up-type partners, singling out the optimal ways to search for this type of new physics, and also give the direction for further model-building (e.g. including other resonances) driven by a need to generate viable parameters of the up-parters sector or induce additional sizable effects needed to compensate the unwanted contributions of the up-type-partners.   
Other composite fermionic resonances are expected to be coupled in a much weaker way compared to those mentioned above and are not restricted to be close to the electroweak scale. Thus they naturally induce smaller distortions of the observables with respect to the SM, which can be further minimized by using a larger freedom in choosing their masses and couplings.
We will nevertheless comment on the effects where the unrelated to naturalness sectors of the model can have large impact on the observables.

In the next sections we will first discuss the PNGB Higgs alone, then add a detailed description of the top quark sector, and eventually extend our framework  to the first two families of quarks. 

\subsection{PNGB Higgs}
\label{sec:ccwz}

We start by introducing the key elements needed to describe the Higgs as a pseudo Nambu-Goldstone boson. They are provided by the general CCWZ formalism~\cite{ccwz}, once one specifies the global symmetry breaking pattern, for which we choose the minimal $SO(5)\times U(1)_X\to SO(4)\times U(1)_X$~\cite{Agashe:2004rs}. 
The spontaneous $SO(5) \to SO(4)$ breaking in the strong sector produces four massless Goldstone bosons, forming an SM-like Higgs doublet. At the same time this scheme prevents the strong sector from breaking the custodial symmetry. An additional $U(1)_X$ factor is introduced in order to reproduce the correct SM hypercharge $Y=T_R^3+X$. 
The Goldstone bosons enter the Lagrangian in a form of a Goldstone matrix
\bea
U=\exp\left[i \frac{\sqrt{2}}{f_{\pi}} \Pi^{\hat a} T^{\hat a}\right]=\,
& \left(\bry{ccc|cc}
& & \vspace{-3mm}& & \\
 & \mathbb{I}_{3} & & &  \\
  & & \vspace{-3mm}& &  \\ \hline
   & & & \cos \frac{\langle h \rangle + h}{f_{\pi}} & \sin \frac{\langle h \rangle + h}{f_{\pi}} \\
    & & & -\sin \frac{\langle h \rangle + h}{f_{\pi}} & \cos \frac{\langle h \rangle + h}{f_{\pi}} \ery\right)\,,
\eea
where $T^{\hat a}$ are the $SO(5)/SO(4)$ generators, $\Pi^{\hat a}$ -- Goldstone bosons and $f_{\pi}$ -- Goldstone decay constant. 
We use the following convention for $\textrm{SO}(5)$ generators
\bea
(T^a_{L,R})_{IJ} = -\frac{i}{2}\left[\frac{1}{2}\varepsilon^{a b c}
\left(\delta_I^b \delta_J^c - \delta_J^b \delta_I^c\right) \pm
\left(\delta_I^a \delta_J^4 - \delta_J^a \delta_I^4\right)\right]\,, \;\;\;\;\;
\label{eq:SO4_gen} 
T^{\hat a}_{IJ} = -\frac{i}{\sqrt{2}}\left(\delta_I^{\hat a} \delta_J^5 - \delta_J^{\hat a} \delta_I^5\right)\,,
\label{eq:SO5/SO4_gen}
\eea
where $T^a_{L,R}$ ($a = 1,2,3$) are the $\textrm{SO}(4) \simeq \textrm{SU}(2)_L \times \textrm{SU}(2)_R$ unbroken generators. For them we will also use an equivalent notation $T^{a}$ with $a = 1\ldots6$. 

We will work in the basis where the gauged ${\cal G}_{\text{SM}}=SU(2)_L \times U(1)_Y$ symmetry of the elementary sector, external to the composite one, is embedded into the unbroken $SO(4)\times U(1)_X$. Such that  the $SU(2)_L$ and $U(1)_Y$ SM bosons gauge the $SU(2)_L$ factor of $SO(4)$ and the $(T^3_R+X)$ generators respectively.
In this basis the non-zero Higgs boson VEV breaks ${\cal G}_{\text{SM}}$.  
The ratio of the Higgs VEV $\langle h \rangle \simeq v=246$~GeV and the $SO(5)$ breaking scale $f_{\pi}$ defines the degree of tuning of the scalar potential~\cite{Agashe:2004rs}
\be
\xi=\left({v \over f_{\pi}}\right)^2\,
\ee
since $\langle h \rangle$ generically tends to get close to $f_{\pi}$, unless the parameters responsible for the generation of $\langle h \rangle$ are finely adjusted. The value of $f_{\pi}$ needs to be somewhat large in order to suppressed the new physics effects, but not too far from $v$ to maintain a tolerable tuning. In this work we will test the value $\xi=0.1$.

The Goldstone matrix in the lagrangian acts as a link between external elementary fields transforming in $SO(5)$ and the composite fields transforming as $SO(4)$ multiplets, thus making the theory invariant under non-linearly realized $SO(5)$.
Hence in order to couple the SM gauge fields to the composite resonances we need to ``dress" them with $U$ matrices. To do this we introduce CCWZ $d$ and $e$ symbols, 
\be
-U^t [A_\mu + i \partial_\mu ] U = d_\mu^{\hat a} T^{\hat a} + e^{a}_{\mu} T^a 
+ e^{X}_{\mu} \,,
\label{eq:d_e}
\ee
where 
$A_{\mu}$ stands for ${\cal G}_{\text{SM}}$ gauge fields 
\bea
A_\mu &=& \frac{g}{\sqrt{2}}W^+_\mu\left(T_L^1+i T_L^2\right)+\frac{g}{\sqrt{2}}W^-_\mu\left(T_L^1-i T_L^2\right)+ \nn\\
&&g \left(\cw Z_\mu+\sw A_\mu \right)T_L^3+g' \left(\cw A_\mu-\sw Z_\mu \right)(T_R^3+Q_X)\,,
\label{gfd}
\eea
where $\cw$ and $\sw$ are the cosine and the sine of the weak mixing angle, $g$, $g'$ are the SM 
couplings of $\textrm{SU}(2)_L$ and $\textrm{U}(1)_Y$ respectively and $Q_X$ is the $X$-charge matrix.
Expanding the definitions~(\ref{eq:d_e}) in fields we have
\bea
d_\mu^{\hat a}&=&\frac{\sqrt{2}}{f_{\pi}}(D_\mu h)^{\hat a}+{\cal O} (h^3) \,,\\
e^{a}_\mu&=&-A^{a}_\mu-\frac{i}{f_{\pi}^2}(h{\lra{D}_\mu}h)^a+{\cal O} (h^4) \label{eq:e_exp} \,,\\
e^{X}_\mu&=&-g^{\prime}  Q_X B_\mu \,,
\eea
where $B_{\mu}$ is the $U(1)_Y$ gauge boson.
%
%
Using $e$ symbols one can construct covariant derivatives acting on the composite sector fields, for instance for the $\Psi$ field transforming in the fundamental representation of $SO(4)$ we have
\begin{equation}
\nabla_\mu\Psi \,=\,\partial_\mu\Psi+i\,e_{\mu}^at^a\Psi\,.
\label{covder}
\end{equation}
It is also convenient to define the analogs of the field strength tensors
\bea
e_{\mu \nu} &=& \partial_{\mu} e_{\nu} - \partial_{\nu} e_{\mu} + i g_{\rho} [e_{\mu},e_{\nu}] \,,\\
e_{\mu \nu}^{X} &=& \partial_{\mu} e^X_{\nu} - \partial_{\nu} e^X_{\mu} \,.
\eea

\subsection{Fermions in \FLFR}
\label{sec:EFT_five}

In this Section we specify in more details the lagrangian of the top partners in the \FLFR model. 
The composite operator ${\cal O}_q$ of the UV mixing lagrangian~(\ref{eq:pc}) can excite a fourplet $\psi_4$ and a singlet $\psi_1$ of the unbroken $SO(4)$ symmetry according to the decomposition ${\bf 5}_{SO(5)}={\bf 4}_{SO(4)}+{\bf 1}_{SO(4)}$. In the following we will also use a notation $\psi=\psi_1+\psi_4$. Using the CCWZ covariant derivative~(\ref{covder}) and the $d$-symbol~(\ref{eq:d_e}) we can write down a lagrangian for the lowest level of composite fermionic excitations
\begin{equation}
{\cal L}_{comp} = i\overline \psi_4 \slashed{\nabla} \psi_4 + i \overline \psi_1 \slashed{\nabla} \psi_1+ \frac{f_{\pi}^2}{4}d_\mu^id^{\mu,i}
- m_4 \overline \psi_4 \psi_4 - m_1 \overline \psi_1 \psi_1
+ \left(i\, c_{41}\, \overline \psi_4^i \gamma^\mu d_\mu^i \psi_1 + {\rm h.c.}\right)\,,
\label{eq:lag_5_5_comp}
\end{equation}
where the $d^2$ term contains the kinetic term of the Goldstone bosons. 
For simplicity we omitted the couplings to gluons, they are trivially deduced from the fact that the top partners must be color triplets in order to mix with the top. We also imposed the parity symmetry in the strong sector, which made the $d$-symbol interactions insensitive to the fermion chirality.
In order to fix the $X$-charge of the top partners we must specify the properties of the top quark.
The SM states clearly do not form the complete $SO(5)\times U(1)_X$ representations, but are embedded into {\bf 5} according to their transformation properties under ${\cal G}_{\text{SM}}$
\begin{equation}\label{eq:embedding_5+5}
q_L^{\bf 5} = \frac{1}{\sqrt{2}} \left(
\begin{array}{c}
i\, b_L\\
b_L\\
i\, t_L\\
-t_L\\
0
\end{array}
\right)\,,
\qquad \quad
t_R^{\bf 5} = \left(
\begin{array}{c}
0\\
0\\
0\\
0\\
t_R
\end{array}
\right)\,,
\end{equation}
where both $q_L$ and $t_R$ embeddings have the same $X$-charge $2/3$, allowing to reproduce the correct electric charge of the top.  
The $q_L=\{t_L,b_L\}$ has an isospin $T_R^3=-1/2$ which provides a protection from large deformations of the $b_L$ couplings~\cite{P_LR}.
The composite operator ${\cal O}_q$ and the top partners hence also have the $X$-charge equal to 2/3 and we can write down the decomposition of the fourplet in terms of $T^3_{L,R}$ eigenstates as
\begin{equation}\label{eq:4-plet_structure}
\psi_4 = \frac{1}{\sqrt{2}}
\left(
\begin{array}{c}
i B - i\, X_{5/3}\\
B + X_{5/3}\\
i\,T + i\,X_{2/3}\\
-T + X_{2/3}
\end{array}
\right)\;,
\end{equation}
wehere $\{T,B\}$ has the left-handed SM doublet quantum numbers, while the $SU(2)_L$ doublet $\{\Xft,\Xtt\}$ has larger by one unit electric charges. 

The elementary part of the lagrangian is trivial and contains the standard covariant derivatives of the elementary quarks. The mixing between the elementary and composite states can be written as
\bea
{\cal L}_{mix} &=& y_{L4} f_{\pi} \left(\overline q_L^{\bf 5} U\right)_i \psi_4^i
+ y_{L1} f_{\pi} \left(\overline q_L^{\bf 5} U\right)_5 \psi_1 + {\rm h.c.}\nonumber\\
&& +\, y_{R4} f_{\pi} \left(\overline t_R^{\bf 5} U\right)_i \psi_4^i
+ y_{R1} f_{\pi} \left(\overline t_R^{\bf 5} U\right)_5 \psi_1 + {\rm h.c.}\,,
\label{eq:lag_5_5_mix}
\eea
where the Goldstone matrices $U$ were introduced as compensators to provide non-linearly realized $SO(5)$ symmetry.
This lagrangian preserves $\cal{G}_{\text{SM}}$ if $\langle h \rangle=0$ ($\langle U \rangle =1$) but breaks the Goldstone symmetry given that the quark embeddings do not form the complete multiplets.
In the following we will often consider the configurations with $y_{L1}=y_{L4}=y_L$ and $y_{R1}=y_{R4}=y_R$. They naturally arise in deconstructed models of CH~\cite{Panico:2011pw,DeCurtis:2011yx} and allow to decrease the sensitivity of the Higgs potential to the cutoff scale. 
The mixings $y$ are expected to be small, realizing weak breaking of the Goldstone symmetry and hence a sufficiently low Higgs mass. The masses of the composite resonances are restricted to $\lesssim 2$~TeV range by the naturalness considerations. The size of the $d$-symbol coefficient $c_{41}$ is expected to be of the order one by the power counting~\cite{Giudice:2007fh}.

On top of the leading order interactions described above, certain observables can be significantly affected by the higher dimensional operators, suppressed by the cutoff scale of our effective description. In order to estimate these effects we will adopt the power counting rule of Ref.~\cite{Giudice:2007fh}, which assumes that the UV effects are characterised by a mass scale $m_{\rho}$ and a coupling $g_{\rho}$, satisfying the relation $m_{\rho} \sim g_{\rho} f_{\pi}$. 
For instance, applying this rule to estimate the size of four-fermion contact interactions we obtain 
\be
{\cal L}_{\text{4ferm}} \sim {1 \over f_{\pi}^2} \psi^4 \,.
\label{eq:4ferm_1}
\ee
The coefficients of the four-fermion operators can be further restricted to specific values, for instance if we assume that they are generated exclusively by the vectorial resonances predicted by some particular UV completion. We will not do this trying to keep the discussion as general as possible and not enter in the additional CH model-building details. Given the presence of a certain degree of freedom in choosing the coefficients of the four-fermion operators, we will try to base our conclusions on the viability of the considered scenarios basing on the least sensitive to them parameters. 

\subsubsection*{Mass spectrum}

The mass spectrum is trivially obtained in the leading order in $v/f_{\pi}$. The top mass is proportional to the left and right mixings with the partners
\be\label{eq:top_mass_5+5}
m_{top}^2 \simeq \frac{\left( y_{L1} y_{R1} m_4 - y_{L4} y_{R4} m_1 \right)^2 f_{\pi}^4}
{(m_4^2 + y_{L4}^2 f_{\pi}^2)(m_1^2 + y_{R1}^2 f_{\pi}^2)}
\frac{\xi}{2} \,,
\ee
the composite states are approximately organized in two $SU(2)_L$ doublets and one singlet $\psi_1=\widetilde T$
\begin{eqnarray}
m_{\Xtt} \simeq m_{\Xft} & = & m_4\,\\
m_{T} \simeq m_B & = & \sqrt{m_4^2 + y_{L4}^2 f_{\pi}^2} \,,\\
m_{\widetilde T} &\simeq & \sqrt{m_1^2 + y_{R1}^2 f_{\pi}^2} \,
\end{eqnarray}

\subsubsection*{Higgs mass}

The Higgs mass arises at one loop level as a consequence of Goldstone symmetry breaking. One unavoidable and large source of this breaking is the mixings of the elementary top with the top partners. These mixings have to be large in order to generate the observed top mass. 
After fixing the mixing parameters $y_{L4}=y_{L1}=y_{L}$ and $y_{R4}=y_{R1}=y_{R}$  the \FLFR model provides a computable~\footnote{The potential in general is logarithmically divergent, but the single divergent operator can be eliminated with a counterterm fixing $\langle h \rangle$.} (UV-insensitive) Higgs mass~\cite{Matsedonskyi:2012ym}
\be
m_h \simeq m_{top} {\sqrt {2 N_c} \over \pi} {m_T m_{\widetilde T} \over f_{\pi}} \sqrt{ \log (m_T / m_{\widetilde T}) \over m_T^2 - m_{\widetilde T}^2}
\label{eq:mh_2s}
\ee
One should however keep in mind that the \FLFR model is supposed to contain only the low-energy (and minimally required) part of some more complete one. For instance, extending it to a three-site model of Ref.~\cite{Panico:2011pw} brings non-negligible changes to the dependence $m_{h}(m_{T},m_{\widetilde T})$, especially in the region with $m_T \sim m_{\widetilde T}$, allowing for significantly heavier partners than predicted by Eq.~(\ref{eq:mh_2s}) (up to $2$~TeV for $\xi = 0.1$, see Ref.~\cite{Matsedonskyi:2012ym}).
 
\subsection{Fermions in \FLOR}
\label{sec:EFT_fourteen}

The main differences of the \FLOR model~\cite{Pappadopulo:2013vca} with respect to the \FLFR model are the larger dimensionality of the composite operator mixing with $q_L$ and the fact that $t_R$ belongs to the strong sector.
The operator $\cal O$ can now excite three multiplets of $SO(4)$: {\bf 9}, {\bf 4} and {\bf 1}. We embed the SM fermions in such a way that they have the same quantum numbers as in the previous case
\begin{equation}\label{eq:embedding_14+1}
q_L^{\mathbf{14}} = \frac{1}{\sqrt{2}}
\left(
\begin{array}{ccccc}
0 & 0 & 0 & 0 & i\,b_L\\
0 & 0 & 0 & 0 & b_L\\
0 & 0 & 0 & 0 & i\,t_L\\
0 & 0 & 0 & 0 & -t_L\\
i\, b_L & b_L & i\,t_L & -t_L & 0
\end{array}
\right)\,.
\end{equation}
and hence one needs the same $X$-charge assignment for all the fermions $Q_X=2/3$. We have already discussed {\bf 4} and {\bf 1}, and 
the nineplet can be decomposed in $T_L^3$ and $T_R^3$ eigenstates as
\begin{equation}\label{decomposition}
\psi_9\supset \{ {\Xet,\Xft,\Xtt} \},\, \{ {\Yft,\Ytt,\Yot} \} ,\, \{ {\Ztt,\Zot,\Zft} \},
\end{equation}
separated according to their $T^3_R=+1,0,-1$ eigenvalues, where subscripts correspond to electric charges. The full matrix form of $\psi_9$ can be found for instance in Ref.s~\cite{Matsedonskyi:2014lla,Pappadopulo:2013vca}. 
The leading order composite lagrangian has the form
\bea
{\cal L}_{comp}&=&i\,\bar t_R \,\slashed \nabla \, t_R + i\overline \psi_9 \, \slashed \nabla \, \psi_9 + i\overline \psi_4 \, \slashed \nabla \, \psi_4 + i \overline \psi_1 \, \slashed \nabla \, \psi_1  
+ \frac{f_{\pi}^2}{4}d_\mu^id^{\mu,i}  \nn\\
&& - m_9 \overline \psi_9 \psi_9 - m_4 \overline \psi_4 \psi_4 - m_1 \overline \psi_1 \psi_1 \nn\\
&& + i\, c_{4t}\, \overline \psi_{4R}^i \gamma^\mu d_\mu^i t_R
+ i\, c_{41}\, \overline \psi_4^i \gamma^\mu d_\mu^i \psi_1 
+ i\, c_{94}\, \overline \psi_9^{ij} \gamma^\mu d_\mu^i \psi_{4j} + {\rm h.c.} , \label{eq:lagr_comp_14}
\eea
where for the covariant derivative of $\psi_9$ we have 
\be
\overline \psi_9 \slashed \nabla \psi_9 =  \, \bar\psi_9^{i,j}\left( \delta^{j,k} \partial_\mu  - 2/3 i g' B_\mu \delta^{j,k} +2i\,\slashed{e}^aT_a^{j,k}\right)\psi_9^{k,i} .
\ee
Given that in a presence of chiral composite state $t_R$ one can not impose a parity in the composite sector, all the terms in the lagrangian~(\ref{eq:lagr_comp_14}) with $d$-symbols can be split in two, with independent coefficients, which we do not do for simplicity. We also omit the mixing term between $\psi_1$ and $t_R$ for the same reason.  
The mixing lagrangian now contains the direct mass term between the elementary $q_L$ and composite $t_R$
\bea
{\cal L}_{mix}  & = & \frac{y_{Lt}}{2} f_{\pi} (U^t \overline q_L^{\bf 14} U)_{55} t_R   \\
&&+ y_{L4} f_{\pi} (U^t \overline q_L^{\bf 14} U)_{i5} \psi_4^i + \frac{y_{L1}}{2} f_{\pi} (U^t \overline q_L^{\bf 14} U)_{55} \psi_1 \\
&&+ y_{L9} f_{\pi} \, (U^t \overline q_L^{\bf 14} U)_{j,i} \,\psi_9^{i,j}+ \textrm{h.c.}  
\eea

\subsubsection*{Mass spectrum}

The mass spectrum is simple to obtain, the top mass is controlled by the $y_{Lt}$ mixing
\be
m_{top}^2 \simeq  \frac{m_4^2}{m_4^2 + y_{L4}^2 f_{\pi}^2} y_{Lt}^2 f_{\pi}^2 \frac{\xi}{2} ,
\ee
the composite resonances in {\bf 4}+{\bf 1} have masses~\cite{DeSimone:2012fs}
\bea
m_{\Xtt} = m_{\Xft} & = & m_4\,,\\
m_T  \simeq m_B &\simeq& \sqrt{m_4^2 + y_{L4}^2 f_{\pi}^2} \,,\\
m_{\widetilde T} &\simeq& m_1  \,,
\eea
while the masses of the members of {\bf 9} are almost degenerate
\bea
m_{\psi_9} &\simeq & m_9\,.
\eea

\subsubsection*{Higgs mass}

Though the \FLOR model does not provide the calculable Higgs mass, the latter can be estimated approximately~\cite{Panico:2012uw,Pappadopulo:2013vca}. 
This estimate shows that, in order to minimize the tuning, the composite fermionic resonances saturating the Higgs potential should not be heavier than $1-2$~TeV.

\subsection{Flavour Patterns}

In Sections~\ref{sec:EFT_five} and~\ref{sec:EFT_fourteen} we have introduced two models of composite partners of the top quark. In this Section we will describe their possible generalizations needed to incorporate the first two families of the up-type quarks. 
Among the considered so far types of CH flavour patterns one can single out flavour-anarchic and flavour-symmetric scenarios. 
The main phenomenological constraints on the anarchic scenarios follow directly from the intrinsic effects of the strong dynamics and Partial Compositeness. Hence we do not expect that accounting for the subtle implications of PNGB nature of the Higgs, which mostly affects EWSB-related observables, can significantly modify the understanding of this scenario. Nevertheless some of the effects discussed in this work should also be accounted for when performing a rigorous analysis of the anarchic scenarios.  
In this work we will focus on flavour-symmetric scenarios, with two types of horizontal symmetries of SM up-type quarks and composite resonances, $U(3)^2$ and $U(2)^2$, the latter acting on the first two families.  Breaking of the given flavour groups will be generated by the interactions of elementary fermions with the composite sector. In order to generate all the flavour structures of the SM it is enough to assign the flavour-breaking couplings to just one chirality of SM quarks, $q_L$ or $u_R$. Corresponding scenarios will be called Right Compositeness (RC) and Left Compositeness (LC). Of course we will assume that the down-type sector also contains some flavour-breaking sources. Their effect will be reflected in the non-diagonal rotation matrices of the down-type quarks and their masses.

\paragraph{ }

\paragraph{\bf \FLFR $\bf U(3)^2$}
\paragraph{ }

We start by considering the case in which the strong sector is symmetric under the diagonal combination of the elementary sector flavour symmetries,  $[U(3)_q \times U(3)_u]_V$, introduced (with an extension to the down partners) in Ref.s~\cite{Cacciapaglia:2007fw,Barbieri:2008zt,Redi:2011zi} in order to minimize the number of flavour-breaking structures. We refer to the extended $[U(3)_q \times U(3)_u \times U(3)_d]_V$-like symmetry as the simplest principle for completing our model with the down-type sector. In this case constraints considered in the present work will also play the dominant role. 


The extension of the one-generation \FLFR model to $U(3)^2$ case is straightforward, one needs to triple the spectrum of composite resonances and promote the mixing terms and mass parameters to three-dimensional matrices in flavour space. The $[U(3)_q \times U(3)_u]_V$ requires all the composite sector parameters to be proportional to the identity in flavour space. 
Here and for the other considered flavour patterns we will also impose a $CP$-conservation in the composite sector.
The general form of the mixing Lagrangian, with the explicitly shown flavour indices, is
\be
{\cal L}_{\text{mass}} = \bar q_{L}^i y_{L}^{i j} f_{\pi}  U \psi^j 
+ \bar u_R^i y_{R}^{i j} f_{\pi}  U \psi^j - \bar \psi^{i} m_{\psi}^{i j} \psi^j \,,
\label{eq:u3mass55}
\ee
where here and in all the following cases we assume $y_{L1}=y_{L4}=y_L$ and $y_{R1}=y_{R4}=y_R$. 
Integrating out composite resonances amounts for a substitution
\be
\psi_{L} \to s_{L}^{\dagger} q_L \, , \;\;\;\;\;\;\; \psi_{R} \to s_{R}^{\dagger} u_R \,,
\ee
where the degree of mixing of the elementary quarks with composites is defined, neglecting EWSB effects, by 
%
%
\be
s_{L}\simeq y_L f_{\pi}  m_4^{-1} { 1\over \sqrt{1+y_{L}^{\dagger} m_{4}^{-2} y_L \, f_{\pi}^2}} \, , \;\;\;\;\;\;\;\; 
s_{R}\simeq y_R f_{\pi}  m_1^{-1} { 1\over \sqrt{1+y_{R}^{\dagger} m_{1}^{-2} y_R \, f_{\pi}^2}} \,.
\label{eq:sinLR_U3}
\ee
After integrating out heavy fermions, we obtain an expression for SM quark masses in the leading order in $v/f_{\pi}$
\bea
m_u & = & s_{L} [m_{1}-m_{4}] s_{R}^{\dagger} \, {v \over \sqrt 2 f_{\pi}}\,,  \label{eq:upmass_u3}\\
m_d & = &\lambda_d \, {v \over \sqrt 2}\,,
\eea
where for the down sector we just assumed a presence of Yukawa interactions without specifying their origin.  In the LC case the matrix $y_L$ (and $s_L$) is proportional to the identity while $y_R$ (and $s_R$) is responsible for flavour breaking. In the RC case the roles of $y_L$ and $y_R$ are switched.    
The diagonalized SM mass matrices and their eigenstates are obtained by $U_{u(d)}$ and $V_{u(d)}$ rotations
\bea
&\hat m_{u(d)}  =  U_{u(d)} m_{u(d)} V_{u(d)}^{\dagger} \;\;\;\;\\
&\begin{cases}
\hat u_L  = U_{u} u_{L}\,, & \hat u_R  =  V_{u} \, u_R \\
\hat d_L  = U_{d} d_{L}\,, & \hat d_R  =  V_{d} \, d_R  \label{eq:mass_basis} 
\end{cases}
\eea
where the symbols with hats correspond to the mass eigenstate basis. We define the matrix $V_{CKM}$ as 
\be
V_{CKM} = U_u U_d^{\dagger}.
\label{eq:v_ckm}
\ee
For the following it is also useful to define the matrix 
\be
\xi_{ij}=V_{\text{CKM} i 3}^{\dagger} V_{\text{CKM} 3 j}
\ee
We take the CKM matrix structure (not necessarily exactly equal to the SM one) and the mass eigenvalues as an input, without trying to explain them.

Given that we assume one of the mixings to be diagonal, all its components should be sizeable in order to generate the top mass, while the second mixing, after diagonalization, has only one large component. 
Therefore for LC the spectrum of composite states contains three degenerate in mass $\{T,B\}$ doublets with a mass $\sim \sqrt{m_4^2+y_{L}^2 f_{\pi}^2}$, one singlet top partner with a mass $\sim \sqrt{m_1^2+\hat y_{tR}^2 f_{\pi}^2}$ and two lighter singlets with masses $\sim m_1$. 
In RC case the spectrum contains three degenerate singlets with mass $\sim \sqrt{m_1^2+y_{R}^2 f_{\pi}^2}$, one top partner doublet $\{T,B\}$ of a mass $\sim \sqrt{m_4^2+\hat y_{tL}^2 f_{\pi}^2}$ and two lighter doublets with masses $\sim m_4$. In addition in all the cases there are three degenerate exotic doublets $\{\Xft,\Xtt \}$ with a mass $\sim m_4$.

Presence of three large elementary-composite mixings also affects the Higgs potential. But as was shown in Ref.~\cite{Matsedonskyi:2012ym}, the resulting expression for the Higgs mass of the two-site model, such as \FLFR, depends dominantly on the product of the left and right-handed mixings. Therefore the Higgs mass is still mostly determined by the top sector and the relation~(\ref{eq:mh_2s}) approximately holds.

\paragraph{\bf \FLFR $\bf U(2)^2$}
\paragraph{ }

A smaller flavour symmetry, still allowing to decrease the number of unwanted flavour-breaking parameters~\cite{Barbieri:2012uh,Redi:2012uj}, is the minimally broken $U(2)^3$~\cite{Barbieri:2011ci}, acting on the first two families of fermions. Again we will only consider the phenomenology of the up-type sector, carrying only the $U(2)^2$ factor. However for concreteness we will also use a specific explicit form of the down-type Yukawa matrices, which can be naturally explained by $U(2)^3$ symmetry.  
 
Implementation of $U(2)^2$ symmetry into \FLFR model does not necessarily require to increase the number of composite partners.  
The dominant contribution to Yukawa interactions of the light quarks could be Technicolor(TC)-like~\cite{Galloway:2010bp}~\footnote{See~\cite{Parolini:2014rza} for the PNGB Higgs with TC-like masses for all the fermions.}, arising from operators bilinear in elementary quarks
\be
{\cal L}_{\text {TC}}^{\text{UV}} \sim  \bar q^{\bf5}_L {\cal O}  u_R^{\bf5} \,,
\label{eq:lag_tc_1}
\ee
where ${\cal O}$ now is a scalar operator transforming in the $SO(5)$ representation. This interaction produces the mass term of the form
\be
{\cal L}_{\text {TC}} = \lambda  \bar q^{\bf5}_L U {\cal S} U^{\dagger}  u_R^{\bf5} \supset \lambda {v \over \sqrt 2} \bar q_L u_R\,,
\label{eq:lag_tc_2}
\ee
where $\cal S$ is some scalar operator with $SO(4)$-invariant VEV $\langle S \rangle \sim f_{\pi}$.
In both cases, assuming partial compositeness or not for light quarks, we are now in principle allowed to embed them differently compared to the third family. For simplicity we will not do this for the \FLFR model.

For both partially composite (PC) and the mixed (called TC in the following) scenarios we assume the following form of the resulting Yukawa matrices, containing the minimal number of flavour breaking needed to generate the SM flavour
\be
\lambda = \left(\begin{array}{ccc|cc}
 & & \vspace{-3mm}& & \\
 & \lambda_{1,2} & & \bf V &  \\
 & & \vspace{-3mm}& &  \\ \hline
 & & & \lambda_3 
\end{array}\right), \;\;\;\;\;\;\;\;
{\cal L}_{\text{mass}} = - {v \over \sqrt 2} \bar u_L^i \lambda^{ij} u_R^j \,,
\label{eq:yukawaU2} 
\ee
where $\lambda_{1,2}$ is a $2\times 2$ matrix and $\bf V$ parametrizes mixing of first two generations with the third one. 
The $\lambda$ matrix of Eq.~(\ref{eq:yukawaU2}) can be approximately diagonalized by left-handed rotation only. We do not give more details on the properties of the Yukawa matrix~(\ref{eq:yukawaU2}), for them we refer the reader to Ref.~\cite{Barbieri:2011ci}. 
For definiteness we also assume the analogous form of the Yukawa matrix in the down sector.

In case of a simple extension of the partial compositeness paradigm to the first two families, the non-diagonal elements of $\lambda$ are generated from $y_L$ and $y_R$ mixings for Right and Left Compositeness respectively. In TC scenarios we could assign some of the flavour breaking sources to the TC interactions, and the others to the mixings with the top partners. For simplicity we will only consider the case when the mixings with the top partners preserve the $U(2)^2$. 
Following these conventions, we present the mixings of the elementary quarks with the composite partners in terms of the Yukawa matrix~(\ref{eq:yukawaU2}) in Table~\ref{tab:sL_sR_u2_55}, where we used the notations
$\Delta_{g_{\psi}} = (m_4-m_1)/f_{\pi}$ with $m_{1,4} =\text{diag}\{ m_{1,4}^u,m_{1,4}^u,m_{1,4}^t \}$. Notice that now even if the light quark partners are present, their mass is not linked to the top partners mass and hence is not restricted to be light. 
\begin{table}
\centering
\begin{tabular}{c | c  | c }
\rule[-6pt]{0pt}{1.5em}                   & $s_L$ & $s_R^{\dagger}$ \\
\hline
\rule[-6pt]{0pt}{1.5em}  ${\bf U(2)^2_{\text{\bf LC}}}$  & $\text{diag}\{ s_{uL},s_{uL},s_{tL} \}$ & $
\Delta_{g_{\psi}}^{-1}
s_L^{-1}
\left(\begin{array}{ccc|cc}
 & & \vspace{-3mm}& & \\
 & \lambda_{1,2} & & \bf V &  \\
 & & \vspace{-3mm}& &  \\ \hline
 & & & \lambda_3 
\end{array}\right)$\\
\rule[-6pt]{0pt}{1.5em}  ${\bf U(2)^2_{\text{\bf RC}}}$ & $\left(\begin{array}{ccc|cc}
 & & \vspace{-3mm}& & \\
 & \lambda_{1,2} & & \bf V &  \\
 & & \vspace{-3mm}& &  \\ \hline
 & & & \lambda_3 
\end{array}\right)
(s_R^{\dagger})^{-1} \Delta_{g_{\psi}}^{-1}$ & $\text{diag}\{s_{uR},s_{uR},s_{tR}\}^{\star}$\\
\rule[-6pt]{0pt}{1.5em}  ${\bf U(2)^2_{\text{\bf TC}}}$ & $
\text{diag}\{0,0,\lambda_{3}/ (s_{tR}^{\star} \Delta_{g_{\psi}})\}$& $\text{diag}\{0,0,s_{tR} \}^{\star}$\\
\end{tabular}
\caption{Mixings of the elementary fermions with composite partners in the \FLFR model with $U(2)^2$ symmetry.}
\label{tab:sL_sR_u2_55}
\end{table}

\paragraph{\bf \FLOR $\bf U(3)^2$}
\paragraph{ }

Extending the \FLOR model to $U(3)^2$ symmetric case encounters a conceptual problem. Since the composite sector is assumed to be flavour-invariant, all the three generations of right-handed up-type quarks must be totally composite. This possibility is ruled out by severe constraints on the degree of compositeness of the first generation of quarks. 


\paragraph{\bf \FLOR $\bf U(2)^2$}
\paragraph{ }

Flavour extension of the \FLOR model in $U(2)^2$-symmetric way also allows for two possible ways of light quark mass generation, but this time we will restrict ourselves to the case of partial compositeness for all the quark families. We will assume \FLOR embedding for the third family of fermions and {\bf {14$_L$+14$_R$}} for the first two families (with elementary $c_R$ and $u_R$), to avoid the constraints on the light quark compositeness. 
For the resulting Yukawa matrices we assume the same form as in Eq.~(\ref{eq:yukawaU2}). In this case, having the $t_R$ as a totally composite state, one can only reproduce the Yukawa matrix~(\ref{eq:yukawaU2}) under the assumption of the Right Compositeness (flavour violation enters only in the interactions with $q_L$). Degree of mixing of partially composite states is defined by
\be
s_{L} = \text{diag}\{\lambda_{1,2}  (s_R^{\dagger})^{-1} \Delta_{g_{\psi}}^{-1} , s_{tL} \}
 \,, \;\;\;\;\;\; 
s_R^{\dagger} = \text{diag}\{s_{1R},s_{1R}\}^{\star}\,,
\ee
where $s_{tL}$ now is not constrained by the size of the top mass.


\section{Experimental Constraints}
\label{sec:pheno}

\subsection{EWPT}

\begin{figure}
\centering
\includegraphics[width=.4\textwidth]{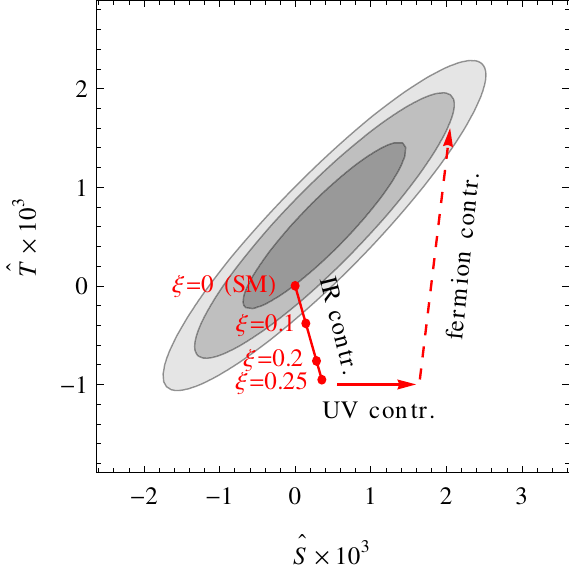}
\caption{Regions in the $\hat S$-$\hat T$ plane, allowed at $68\%$, $95\%$ and $99\%$CL~\cite{Baak:2012kk} (from dark to light grey respectively).
The estimated size of the contributions due to the Higgs couplings modification~(\ref{eq:s_irlog}), composite vectors~(\ref{eq:s_vect}) and composite fermions are shown with red lines. The cutoff scale is taken to be $3$~TeV.}
\label{fig:ST_plane}
\end{figure}

In this Section we will discuss the $\hat S$ and $\hat T$ parameters~\cite{Peskin:S,Barbieri:2004qk} sensitive to the New Physics contributions in the two point functions of electroweak gauge bosons. On Fig.~\ref{fig:ST_plane} we show the currently allowed values of $\hat S$ and $\hat T$  and an estimated size of different corrections to them~\cite{ewpt}, which will be discussed in details in this Section.  Given the approach taken in this work, the central point of our discussion will be the effects related to the up-type partners (for the constraints on the down partners see for instance~\cite{Gillioz:2013pba}).

\subsubsection{${\hat S}$ parameter}
\label{sec:S_param}

The least model dependent~\cite{Barbieri:2007bh} contribution to the $\hat S$ parameters originates from the modification of the Higgs couplings to the gauge bosons and arises at one loop as a logarithmically divergent positive~\cite{Falkowski:2012vh} term
\be
\Delta \hat S^{\text{(Higgs)}} = {g^2 \over 192 \pi^2} \xi \log\left ({m_\rho^2\over m_h^2}\right)\,.
\label{eq:s_irlog}
\ee
The contribution of the UV dynamics to $\hat S$ can be estimated, using the power counting of Ref.~\cite{Giudice:2007fh}
\be
\Delta \hat S^{\text{(vect)}} \simeq {m_{W}^2 \over m_{\rho}^2} \, .
\label{eq:s_vect}
\ee
To consider it in more details, we can assume that it is dominated by an exchange of composite vectorial resonances, transforming in the adjoint representation of $SO(4)$ ($\rho_{\mu}$) and $SO(5)/SO(4)$ ($a_{\mu}$). The resulting correction in this case is proportional to the difference of the $\rho_{\mu}$ and $a_{\mu}$ self-energy derivatives~\cite{Contino:2010rs} and gives 
\be
\Delta \hat S \simeq {\xi \over 2} \left[ {g^2 \over g_{\rho}^2} - {g^2 \over g_{a}^2} \right]\,,
\label{eq:s_vect_ax}
\ee
where $g_{\rho}$ and $g_a$ are respectively the couplings of the $\rho_{\mu}$ and $a_{\mu}$ resonances.
Despite the fact that $\Delta \hat S$ is given by a difference, the resulting contribution is typically positive.
The positivity follows from the symmetry properties of the explicit CH models or from the Weinberg sum rules~\cite{wsr} requiring a good behaviour of the two-point functions at high energies. Nevertheless certain non-minimal terms in the Lagrangian, which we will not consider in this work, can lead to an overall negative sign of the vector resonance contribution~\cite{DeCurtis:2014oza}.    

\begin{figure}[t]
\centering
\includegraphics[width=0.49\textwidth]{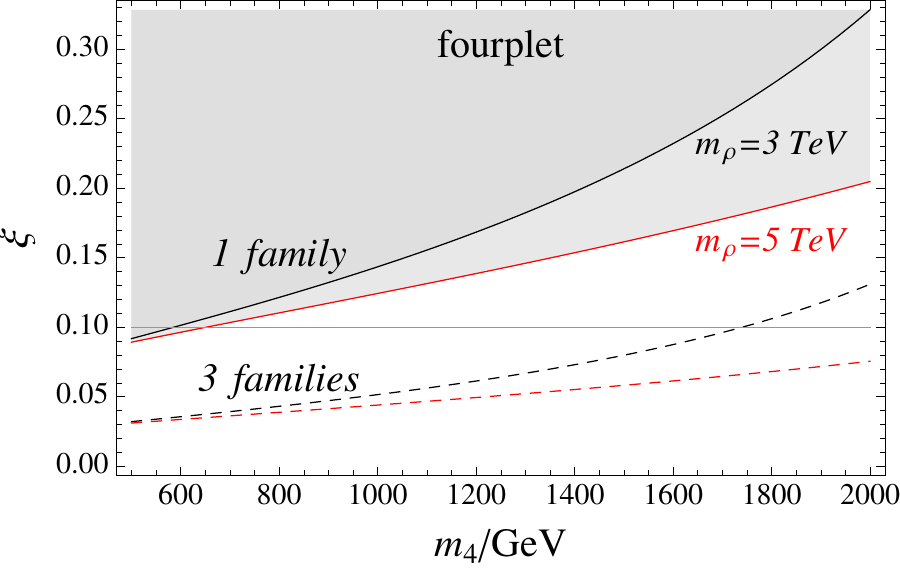}
\hfill
\includegraphics[width=0.49\textwidth]{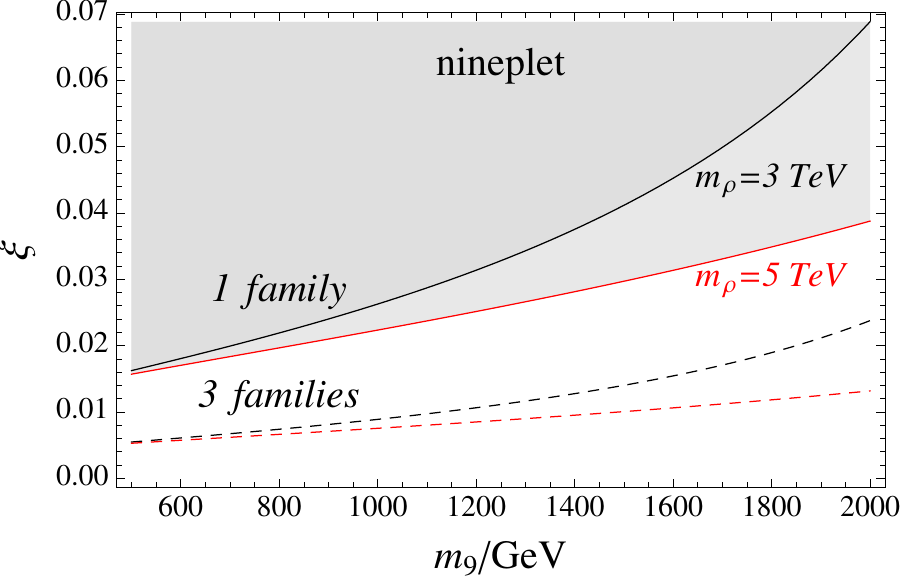}
\caption{99\%CL bounds on $\xi$ obtained from the $\hat S$ parameter only, for a cutoff scale $3$~TeV (black lines) and $5$~TeV (red lines), for one family of composite resonances (solid lines) and three degenerate families (dashed lines), for the resonances transforming as {\bf 4} (left panel) and {\bf 9} (right panel) of SO(4). The green line bounds the region with a moderate tuning  $\xi\sim0.1$. The $\hat S$ parameter, besides fermionic contribution, includes the IR logarithm (\ref{eq:s_irlog}) and a UV part (\ref{eq:s_vect}).}
\label{fig:S_param_1}
\end{figure} 

As was noticed in Ref.s~\cite{ewpt,Azatov:2013ura,Barbieri:2008zt}, the $\hat S$ parameter can also receive logarithmically divergent contribution from the loops of composite fermions, independent on their mixings with the elementary states. The size of this contribution depends on the fermionic representations, for those relevant for the present discussion we have~\cite{Lavoura:1992np}
\be
\Delta \hat S^{\text{(ferm)}} = {g^2 \xi \over 16 \pi^2} \left[
 2 \, n_{\psi_4} + 12\, n_{\psi_9} -  4\, c_{41}^2\, n_{{\psi_4 d \psi_1}} - 2\, c_{4t}^2 \, n_{{\psi_4 d t_R}} - 9\, c_{94}^2 \, n_{{\psi_9 d \psi_4}}  \right]\log m_{\rho}^2 \, ,
 \label{eq:S_log_all}
\ee
where the first two terms correspond to contributions of  $n_{\psi_4}$ fourplets and $n_{\psi_9}$ nineplets respectively, and the rest corresponds to the $d$-symbol interactions between different multiplets.
Given a potentially large number of relatively light composite fermion families, the contribution~(\ref{eq:S_log_all}) can dominate the $\hat S$. 

This correction can be seen as originated from the running of the effective couplings $g_{\rho}$ and $g_a$ in Eq.~(\ref{eq:s_vect_ax}) from the mass scale of composite vectors down to the electroweak scale, induced by composite fermions\footnote{The running of the $SO(5)/SO(4)$ resonance contribution to $\hat S$ has additional corrections not accounted for just by running of the coupling, and proportional to the composite fermion mass difference. This behaviour is explained by the fact that the coset resonances correspond to the symmetry which is not respected by the strong sector.} 
\be
g_{\rho,a}^{\text{eff}\,2} 
\simeq g_{\rho,a}^2 - \beta_{\rho,a} \log \left(m_{\rho,a}^2 \over {m_{\psi}^2}\right) \;\;\;\; \text{with} \;\;\;\;
\{\beta_{\rho},\beta_{a}\} = N_c N_f C_{\bf r} {g_{\rho,a}^4 \over 12 \pi^2} \cdot \{1, 2 c_{41}^2\}\,,
\label{eq:g_rho_eff}
\ee
where $N_c$ and $N_f$ are the numbers of colors and flavours and $C_{\bf r}$ is a Casimir operator of the fermionic representation. 
The effective $g_{a,\rho}$, evaluated at the electroweak scale, are not subject to the Weinberg sum rules and hence can allow for a negative overall shift in $\hat S$. 

Therefore the $\hat S$ parameter is sensitive both to the cutoff scale and to the composite resonances not interacting with the quark sector.
On Fig.~\ref{fig:S_param_1} we show the bounds on the fourplet and nineplet masses, depending on the number of composite families with degenerate mass parameters, contributing to $\hat S$. Three degenerate families of fourplets or even just one family with a nineplet are severely constrained by $\hat S$, hence in the following, considering these possibilities, we will need to tune the $c$-coefficients to cancel large positive contributions to $\hat S$. 


\subsubsection{${\hat T}$ parameter}

The two dominant sources of corrections to the $\hat T$ parameter in CH models with custodially-symmetric composite sector come from the distortion of the Higgs couplings and from the elementary-composite fermion mixings. The former effect is analogous to the logarithmically divergent contribution to $\hat S$~(\ref{eq:s_irlog}), but is negative and has a larger size
\be
\Delta \hat T^{h} = - {3 g^{\prime\,2} \over 64 \pi^2} \xi \log \left( {m_{\rho}^2 \over m_h^2} \right) \,,
\ee
which for a reference value $m_{\rho} = 3$~TeV becomes $\Delta \hat T^{h} \simeq - 4 \cdot 10^{-3} \xi$.
Therefore typically the fermionic contribution to $\hat T$ needs to be positive to compensate $\Delta \hat T^{h}$. This becomes not strictly necessary if $\hat S$ receives large negative contribution from the fermionic loops, but in the following we will prefer not to consider this possibility since it requires the $d$-symbol coefficients to be slightly larger than what is typically expected.  

In both considered types of elementary fermion embeddings, the $\hat T$ parameter is finite and hence can be reliably computed. The parameter, breaking the custodial symmetry, is $y_L$, therefore all the partners with sizable $y_L$ are relevant for $\Delta \hat T$. We present analytic expressions for the fermionic contributions to the $\hat T$ parameter from the composite partners transforming as {\bf 1}, {\bf 4} and {\bf 9} in case of composite $t_R$, relevant for the \FLOR model, computed using the results of Ref.~\cite{Anastasiou:2009rv}. For completeness we also report  $\Delta \hat T$ induced by {\bf 1} and {\bf 4} with a partially composite $t_R$, computed in previous studies~\cite{Gillioz:2008hs,Barbieri:2007bh,ewpt}. All the expressions are given in the leading order in $\xi$ expansion.

\subsubsection*{\FLFR model}

\begin{itemize}

\item
{\bf singlet} gives a positive contribution to the $\hat T$
\be
\Delta T^{5+5 (1)} = \frac{3\, \xi}{64 \pi^2} \frac{y_{L1}^4 m_1^2 f_{\pi}^2}{(m_1^2 + y_{R1}^2 f_{\pi}^2)^3}
\left\{m_1^2 + 2 y_{R1}^2 f_{\pi}^2 \left[
\log\left(\frac{2 (m_1^2 + y_{R1}^2 f_{\pi}^2)^2}{v^2 y_{L1}^2 y_{R1}^2 f_{\pi}^2}\right) - 1\right]\right\}\,.
\label{eq:T55_1}
\ee

\item
{\bf fourplet} instead contributes negatively. 
Performing additional expansion in mixing parameters to simplify the expression, one gets
\be
\Delta T^{5+5 (4)}= - \frac{\xi}{32 \pi^2} \frac{y_{L4}^4 f_{\pi}^2}{m_4^2}\,.
\label{eq:T55_4}
\ee

\end{itemize}

\subsubsection*{\FLOR model}

\begin{itemize}

\item
{\bf singlet} contribution coincides with the one obtained for the ${\bf 5_{\bf L}+1_{\bf R}}$ model~\cite{ewpt} and is positive
\be
\Delta T^{14+1 (1)} = \frac{3\, \xi}{64 \pi^2} \frac{y_{L1}^2 f_{\pi}^2}{m_1^2}
\left\{y_{L1}^2 + 2 y_{Lt}^2 \left[
\log\left(\frac{m_1^2}{m_{top}^2}\right) - 1\right]\right\}\,.
\label{eq:T141_1}
\ee

\item
{\bf fourplet}-induced correction, expanded in $y f / m_4$, reads
\bea
\Delta T^{14+1 (4)} &=& {\xi \over  32 \pi^2} {y_{L4} f_{\pi}^2 \over m_4^2} \Bigg\{ 
-(17+9 c_{4t}) y_{L4}^3 + 18 \sqrt 2 c_{4t} y_{Lt} y_{L4}^2 +6(1+2c_{4t}^2) y_{Lt}^2   \nn \\
&&+3 y_{Lt}^2 ( y_{L4} - 4 \sqrt 2 c_{4t}y_{Lt} ) \left[\log\left( {m_4^2 \over m_{top}^2}\right) -1 \right] \Bigg\}\,,
\label{eq:T141_4}
\eea
and can be of both signs.

\item
{\bf nineplet} alone gives a positive contribution to $\hat T$, but  the right-handed top induces an additional negative shift, proportional to $\log\left({m_9^2 \over m_t^2}\right)$. The latter contribution is controlled by $y_{Lt} \sim y_{top}$ which can not be made much smaller than 1 for the top partners. The positivity of the resulting contribution 
\be
\Delta T^{14+1 (9)}= {\xi \over  64 \pi^2}  {y_{L9}^2 f_{\pi}^2 \over m_9^2} \left\{19 y_{L9}^2 + 6 y_{Lt}^2 \left[7- 3\log\left( m_9^2 \over m_{top}^2  \right) \right]  \right\}
\label{eq:T141_9}
\ee
hence requires a somewhat large $y_{L9}$.

\end{itemize}

\subsection{Flavour-diagonal $Z$ couplings}

\subsubsection{$Z\bar uu$ (and $W\bar ud$)}
\label{sec:ZttWud}

The most stringently constrained variable, sensitive to the variation of the $Z$ couplings to up-type quarks is the hadronic $Z$ width normalized to the leptonic one. According to Ref.~\cite{Baak:2012kk} we have at $1\sigma$
\be
R_h^{\text{exp}} = 20.767(25)\, , \;\;\;\;  R_h^{\text{SM}} = 20.740(17) \,.
\label{eq:rh_bound}
\ee
Couplings to the left- and right-handed light quarks with the $Z$ can be significantly distorted with respect to the SM for, respectively, Left and Right Compositeness with $U(3)^2$. 
The $Z uu$ coupling turns out to be equal in the leading order to another important parameter, $\delta V_{ud}$~\cite{delAguila:2000aa}.
The latter in the LC case defines a universal factor $(1+\delta V_{ud})$ which rescales all the left-handed $W$-boson couplings. Therefore using the constraint on the CKM matrix unitarity~\cite{Agashe:2014kda} 
we can obtain a bound on $\delta V_{ud}$ (at $1\sigma$), 
\be
|V_{ud}^{\text{eff}}|^2+|V_{us}^{\text{eff}}|^2+|V_{ub}^{\text{eff}}|^2=(1+\delta V_{ud})^2 \sum_x |V_{u x}|^2 = (1+\delta V_{ud})^2= 0.9999(6)\,,
\ee
where $V_{ux}^{\text{eff}}$ are the effective CKM elements generated in CH and $V_{ux}$ are the elements of the unitary matrix defined in Eq.~(\ref{eq:v_ckm}).
The deviation from the universal rescaling can in principle appear in $V_{ub}$, due to a sizable $y_{tR}$, but the overall suppression by $V_{ub}$ makes it irrelevant.



For the Lagrangian parametrized as
\be
{\cal L}_{Z,u} = {g \over \cw} Z_{\mu} \bar t \gamma^{\mu} [(g_{u_L}^{\text{SM}} + \delta g_{u_L})P_L + (g_{u_R}^{\text{SM}} + \delta g_{u_R})P_R] t \,,
\ee
where
\be
g_{u_L}^{\text{SM}} = {1\over 2} - {2 \over 3} \sw^2\, , \;\;\;\;\; g_{u_R}^{\text{SM}} = - {2 \over 3} \sw^2\, ,
\ee
the leading contributions to the discussed couplings in \FLFR model are~\cite{ewpt}  
\bea
\delta g_{u_L}^{5+5} &=& \delta V_{ud}^{5+5} = -\frac{\xi}{4} \frac{f_{\pi}^2}{m_4^2 + y_{L4}^2 f_{\pi}^2}\left[
\left(\frac{m_4 m_1 y_{L1} + y_{L4} y_{R4} y_{R1} f_{\pi}^2}{m_1^2 + y_{R1}^2 f_{\pi}^2} - \sqrt{2} c y_{L4}\right)^2
+ (1 - 2 c^2) y_{L4}^2\right] \;\;\;\;\;\;\;
\label{eq:gtL55}
\\
\delta g_{u_R}^{5+5} &=& \frac{\xi}{4} \frac{f_{\pi}^2}{m_1^2 + y_{R1}^2 f_{\pi}^2}
\left[\left(\frac{m_4 m_1 y_{R4} + y_{L4} y_{L1} y_{R1} f_{\pi}^2}{m_4^2 + y_{L4}^2 f_{\pi}^2}
- \sqrt{2} c y_{R1}\right)^2
- \left(\frac{m_1 y_{R4}}{m_4} - \sqrt{2} c y_{R1}\right)^2
\right]\,
\label{eq:gtR55}
\eea
%
$\delta V_{ud}$ and $\delta g_{u_L}$ put very stringent constraints on LC scenario~\cite{Redi:2011zi,Barbieri:2012tu}, therefore we will comment on a condition needed to minimize them.   
As was pointed out in Ref.~\cite{ewpt}, the interactions of the fermions with the Goldstones of the Lagrangians~(\ref{eq:lag_5_5_comp}) and~(\ref{eq:lag_5_5_mix}) can be localized in the single term
\be
{\cal L}^{c=1/\sqrt 2} \supset - (m_1 - m_4) (\bar \psi U)_5 (U^{\dagger} \psi)_5
\label{eq:lag_c1sqrt2}
\ee
after setting $c_{41}=1/\sqrt 2$, $y_{L1}=y_{L4}=y_L$, $y_{R1}=y_{R4}=y_R$ and making a field redefinition $\psi \to U^{\dagger} \psi$.  Therefore for $m_1$ close to $m_4$ all the New Physics effects sensitive to EWSB are decreased. This configuration corresponds to a restored $SO(5)$ symmetry of the composite sector, hence the Goldstone matrices can be removed from the interactions with fermions. This, however, also leads to a vanishing top mass, see Eq.~(\ref{eq:top_mass_5+5}), since in $U(3)^2$ case all the families of partners share the same $m_1$ and $m_4$. 


Let us now consider a case with $c_{41}=-1/\sqrt 2$. Now, after flipping a sign of the singlet partners of the up and charm $\widetilde T_L^{(1,2)} \to -\widetilde T_L^{(1,2)}$, and redefining $m_{1}^{(1,2)} \to -m_{1}^{(1,2)}$, we arrive to almost the same form of the Lagrangian for the \emph {left-handed fermions} (including their interactions with the right-handed ones) of the \emph {first two generations} as in the $c_{41}=1/\sqrt 2$ case. 
The only coupling, whose form will differ, is the mixing of $\psi_L$ with $u_R$ and $c_R$
\be
y_R \bar u_R U \psi_L \to y_R \bar u_R U (\psi^4_L-\psi^1_L)
\ee
but in case of LC it is negligible. Therefore, as follows from the Eq.~(\ref{eq:lag_c1sqrt2}), we can decrease all the EWSB effects in the left-handed up and charm sector for $m_{1}^{(1,2)}\simeq m_{4}^{(1,2)}$. In particular, $\delta V_{ud}$ and $\delta g_{u_L}$ can be suppressed.
The top mass in this case is proportional to $|m_1^{(3)}-m_4^{(3)}| = |m_1^{(1,2)}+m_4^{(1,2)}|$ and hence can be made sufficiently large.
In agreement with our discussion, for $c_{41}=-1/\sqrt 2$, $y_{L1}=y_{L4}=y_L$ and $y_R\to0$ Eq.~(\ref{eq:gtL55}) simplifies to
\be
\delta g_{u_L}^{5+5} = \delta V_{ud}^{5+5} = -\frac{\xi}{4} \frac{y_{L}^2 f_{\pi}^2}{m_4^2 + y_{L}^2 f_{\pi}^2}\left[
\frac{m_1+m_4}{m_1}\right]^2 \,.
\label{eq:gtL55_simp}
\ee
The discussed condition only minimizes the tree-level contributions, the loop effects can in principle be comparable with the current bound
\be
\delta V_{ud}^{5+5 \text{(1loop)}} \simeq \xi {1 \over 16 \pi^2} s_{L}^2
\ee
but their detailed analysis lays beyond the scope of this work. 

The couplings of the right-handed up-type quarks to the $Z$ are protected by the $P_{LR}$, which is only broken by the $y_L$ mixing. Therefore, as can be explicitly checked from Eq.~(\ref{eq:gtR55}), the $\delta g_{u_R}$ is proportional to the product of left- and right-handed mixings, i.e. $\delta g_{u_R,c_R} \sim \hat \lambda_{u,c}$, which is negligible. 
Bounds on the distortions of the right-handed couplings of the $W$ are too weak and hence irrelevant for our discussion~\cite{Barbieri:2012tu}.

In the \FLOR scenario, which is only defined for $U(2)^2$, the overall scale of the couplings deviation is dictated by the degree of compositeness of left-handed light quarks, which is naturally small.


\subsubsection{$Z\bar dd$}
\label{sec:Zbb}

\begin{figure}
\centering
\includegraphics[width=.41\textwidth]{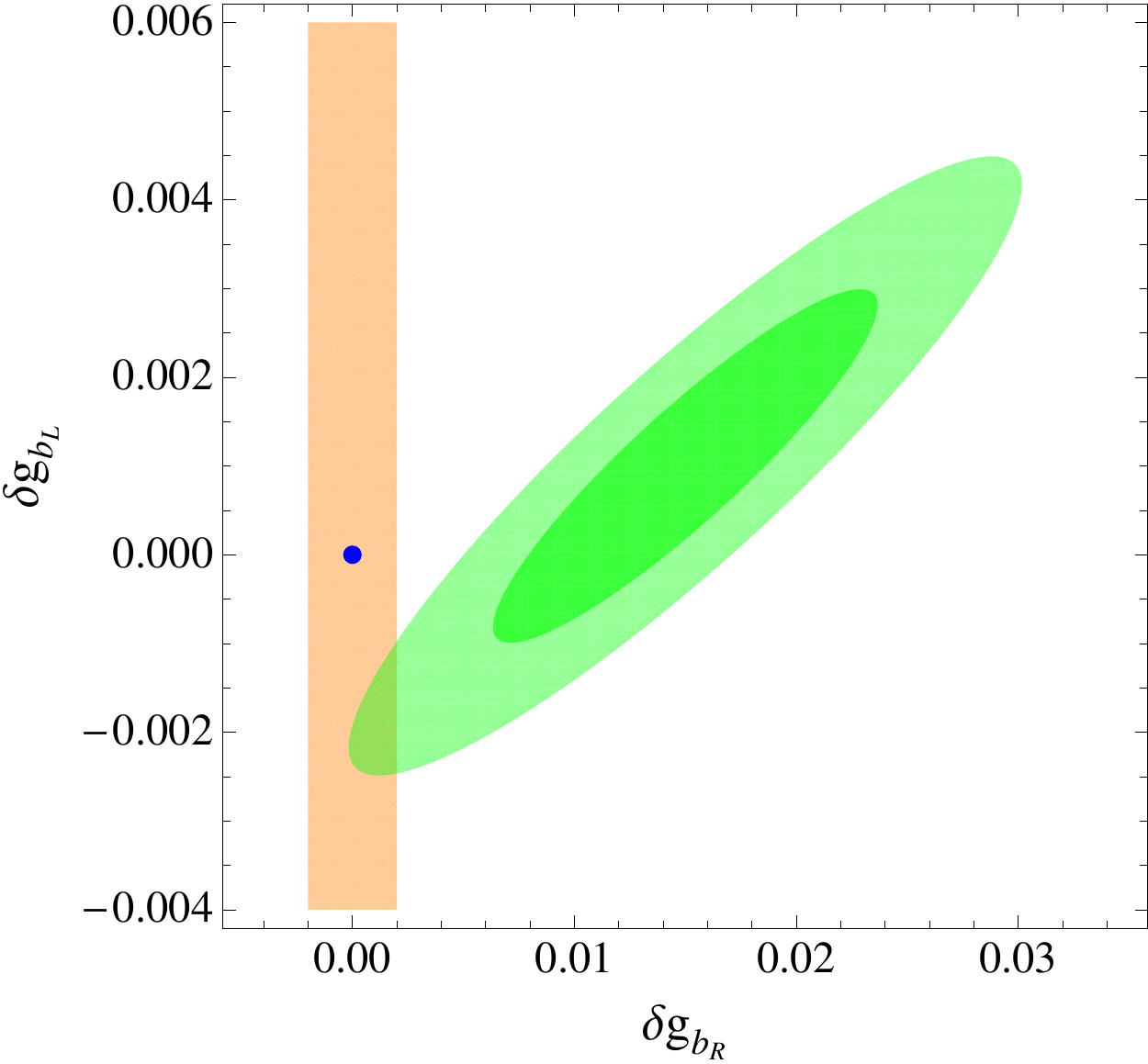}
\hspace{1cm}
\includegraphics[width=.42\textwidth]{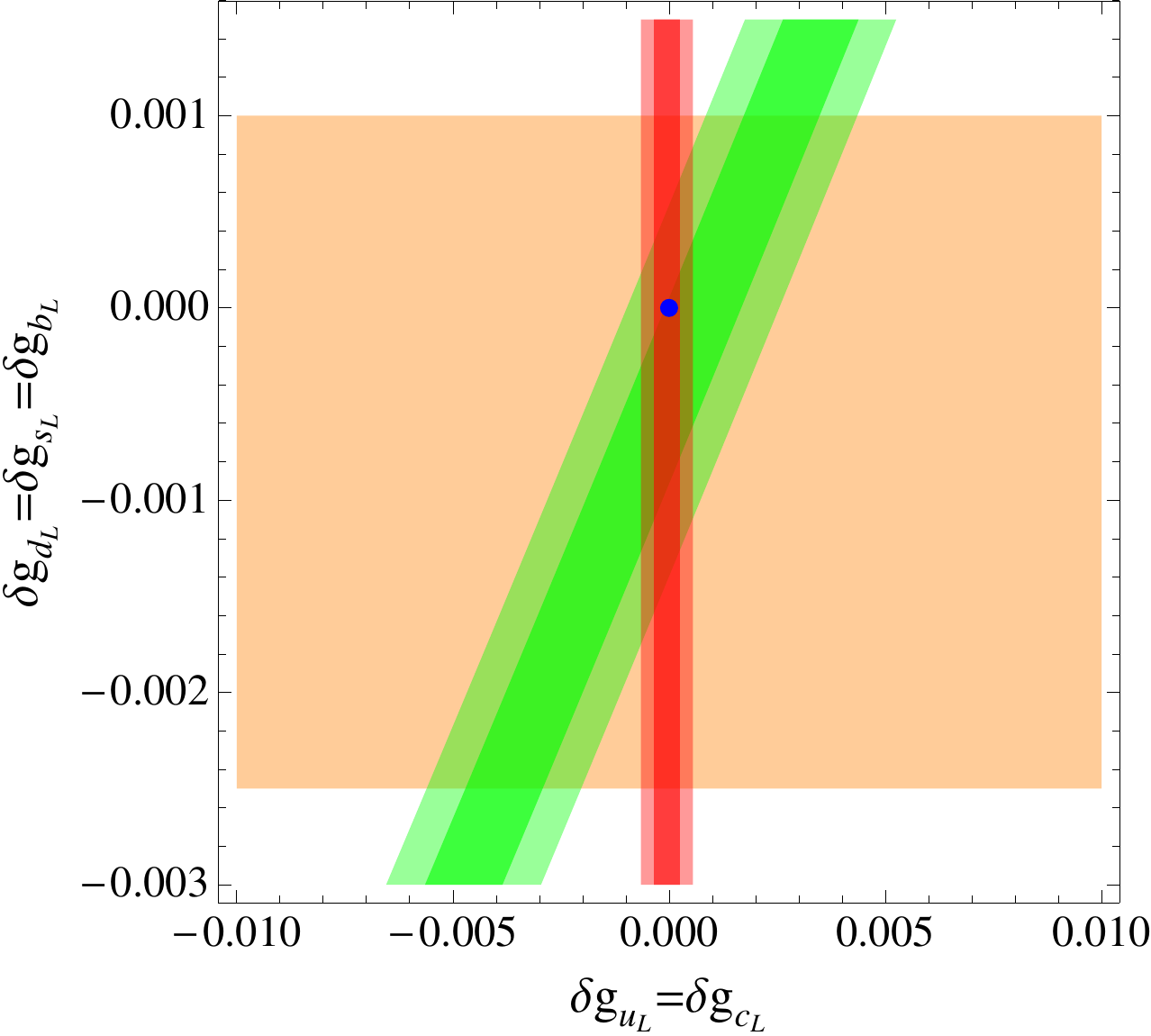}
\caption{Constraints on the corrections to the $Z$ boson couplings to the quarks. Left panel: experimental constraints on $Z \bar bb$ coupling (green ellipses),
at $68\%$ and $95\%$ confidence level~\cite{constraintZbb}; 
the vertical orange band shows the expected size of the corrections to the $g_{b_R}$ coupling.
Right panel: constraints on the left quark couplings in the case of $U(3)^3$ Left Compositeness, derived from $R_h$ at $1\sigma$ and $2\sigma$ (in green); orange band corresponds to the possible size of $\delta g_{b_L}$ compatible with constraints on $Z \bar bb$ coupling. Red bands are allowed by $\delta V_{ud}$(assuming $\delta V_{ud}=\delta g_{u_L}$) at $1\sigma$ and $2\sigma$.  Modifications of the right-handed couplings are taken to be 0.  Blue dots on both plots correspond to the SM predictions.}
\label{fig:deltag_plane}
\end{figure}

In our set-up the $Z\bar bb$ vertex can not receive any tree-level corrections at zero momentum transfer from the top partner sector. 
The reason is the left-right parity $P_{LR}$~\cite{P_LR} in the charge -1/3 sector, protecting $g_{b_L}$, and the absence of interactions with $b_R$.
In order to generate the latter we need to include the bottom partners in our description. Bottom partners can also generate the tree-level corrections to $g_{b_L}$, which are however naturally suppressed by a small mass of the bottom quark.
 %
For the simplest embedding of the $b_R$, as a singlet of $SO(4)$, the right-handed coupling $g_{b_R}$ is also protected  by $P_{LR}$ and hence can only receive one-loop corrections.
Nevertheless, given a disagreement between the current measurement of the $Z\bar bb$ couplings and the SM prediction, it is desirable to have a certain, though small, deviation from the SM, as can be seen on the left panel of Fig.~\ref{fig:deltag_plane}.

In the scenarios with $U(3)$ symmetry also the couplings of the down and strange quarks to the $Z$ can become sizably distorted, modifying the prediction for the hadronic $Z$ width, discussed in the previous Section. On the right panel of Fig.~\ref{fig:deltag_plane} we present the allowed modifications of the left-handed couplings to the up- and down-type quarks, taking $\delta g_{d_L}=\delta g_{s_L}=\delta g_{b_L}$ and $\delta V_{ud}=\delta g_{u_L}=\delta g_{c_L}$, which can be the case for $U(3)^2$ LC.

In the following we will consider the deviations which can be induced by the composite partners. 
We parametrize the momentum-independent $Z\bar dd$ interaction lagrangian as  
\be
{\cal L}^Z = \frac{g}{\cw} Z_\mu \overline d \gamma^\mu \left[
(g^{\text{SM}}_{d_L} + \delta g_{d_L}) P_L + (g^{\text{SM}}_{d_R} + \delta g_{d_R}) P_R\right] d\,,
\ee
where with a one-loop precision
\be
g_{d_L,s_L}^{\text{SM}} = -{1 \over 2} + {1 \over 3} \sw^2 \, , \;\;\;\;\;\; g_{b_L}^{\text{SM}} = -{1 \over 2} + {1 \over 3} \sw^2 + {m_t^2 \over 16 \pi^2 v^2}\,, \;\;\;\;\;\;g_{d_R,s_R,b_R}^{\text{SM}} = {1 \over 3} \sw^2 \,. 
\ee
In the following we will call all the couplings $g_{b_L}$ or $g_{b_R}$, keeping in mind that our results also apply to the first two generations.  

\subsubsection*{Momentum-dependent corrections}


We will first focus on the corrections to the $Z\bar bb$ coupling, proportional to the momentum transfer in the vertex and hence unaffected by the $P_{LR}$ symmetry protection. We will first sketch the general line of the analysis and then go into details. One can write down the following momentum-dependent operator modifying the $Z \bar b_L b_L$ interactions
\be
{\cal L}_{q^2 Zbb} \, \sim \, { g_{b_L}^{\text{SM}} s_L^2 \over m_{\rho}^2} \, \bar b\,  \gamma_{\mu} D_{\nu} F^{\mu \nu} \, b  
\, \to \, g_{b_L}^{\text{SM}} s_L^2 {m_Z^2 \over m_{\rho}^2} \, \bar b\,  \gamma_{\mu} Z^{\mu} \, b  \, \sim \, 1 \cdot 10^{-3} g_{b_L}^{\text{SM}} s_L^2  \, \bar b\,  \gamma_{\mu} Z^{\mu} \, b
\label{eq:p2zbb_simp}
\ee
where two powers of $s_L$ follow from the fact that the operator is generated due to the strong sector dynamics which is only coupled to the bottom by mass mixing. To obtain the second expression we put the $Z$ on its mass shell, for the last estimate we took $m_{\rho}=3$~TeV.  This expression closely resembles the UV correction to the $\hat S$ parameter of Eq.~(\ref{eq:s_vect}). The latter is significantly affected by the one-loop contribution of the fermionic resonances, which can be conveniently estimated by taking $m_{\rho} \sim g_{\rho} f_{\pi} \to g_{\rho}^{\text{eff}} f_{\pi} \sim m_{\rho}^{\text{eff}}$ in Eq.~(\ref{eq:p2zbb_simp}). The $m_{\rho}^{\text{eff}}$, affected by loops with multiple colored resonances, can differ by a factor of few from the initial $m_{\rho}$, potentially enhancing the resulting effect in $\delta g_{b_L}$. Since this effect scales with the $Z$ momentum, it is negligible for the flavour physics observables.

For a more detailed analysis of this effect we will first introduce the bottom partners in a more concrete way, for the purpose of this Section only.
We assume that they form ${\bf 1}_{-1/3}+{\bf 4}_{-1/3}$ under $SO(4)\times U(1)_X$ and mix with the elementary fermions $b_R$ and $q_L^{(b)}$, embedded respectively as the $SO(4)$ singlet and $T_R=+1/2$ component of the fourplet,  with a charge $-1/3$ under $U(1)_X$. The $q_L^{(b)}$-bottom partners mixing induces tree-level corrections to $Z \bar b_L b_L$ and thus has to be very small.  
The $b_L$ and $b_R$ hence are mixed mostly with the top and bottom partners respectively. 
Using the CCWZ building blocks, one can construct the following dimension-six operators with the top and bottom partners contributing to $Z\bar bb$
\bea
{\cal L}_{q^2 Zbb}&=&
{c_{bL}^{\rho} \over m_{\rho}^2}\overline B_L \gamma_{\mu} (\nabla_{\nu} e^{\nu \mu}) B_L + 
{c_{bL}^{X} \over m_{X}^2}\overline B_L \gamma_{\mu} (\nabla_{\nu} e^{(X)\,\nu \mu}) B_L + \nn \\
&&{c_{bR} \over m_{X}^2}\overline{\widetilde B}_R \gamma_{\mu} (\nabla_{\nu} e^{(X)\,\nu \mu}) {\widetilde B}_R
\label{eq:lag_zbb_p2}
\eea
where $B$ is a charge -1/3 top partner and $\widetilde B$ is a singlet partner of $b_R$. 
We kept different mass scales suppressing the different operators, as would be the case if they were generated exclusively by integrating out $\rho$ and $X$ bosons (all the $c$-coefficients are equal to one in this case).  
For the on-shell $Z$ the effect of the couplings of Eq.~(\ref{eq:lag_zbb_p2}) is
\bea
{\delta g_{b_L}^{\text{($q^2$)}}} &=& s_{tL}^2 \,\left\{-{1 \over 2}(1-2 \sw^2)  {m_Z^2 \over m_{\rho}^2} -
 {2\over3} \sw^2   {m_Z^2 \over m_{X}^2} \right\} \simeq -0.4 \cdot 10^{-3} s_{tL}^2   
\label{eq:deltagL}  \\
{\delta g_{b_R}^{\text{($q^2$)}}} &=&  s_{bR}^2 \, \left\{{1\over 3} \sw^2 {m_Z^2 \over m_{X}^2} \right\} \simeq 0.1 \cdot 10^{-3} s_{tL}^2
 \label{eq:deltagR}
\eea
where $s_{bR}$ is the $b_R$ degree of compositeness and the numerical estimates were made assuming $m_{\rho,X}=3$~TeV.
After accounting for one-loop fermionic corrections, the coefficients of the operators~(\ref{eq:lag_zbb_p2}) can be enhanced in analogy with the $\hat S$. 
Notice, however, that the two effects, though having the same origin, are not correlated. Indeed, the one-loop correction to the $\hat S$ is related to the difference between the $SO(4)$ and $SO(5)/SO(4)$ form-factors, while the correction to $Z\bar bb$ comes mostly from a combination of  the $SO(4)$ and $X$-charge ones.

\begin{figure}[t]
\centering
\includegraphics[width=0.15\textwidth]{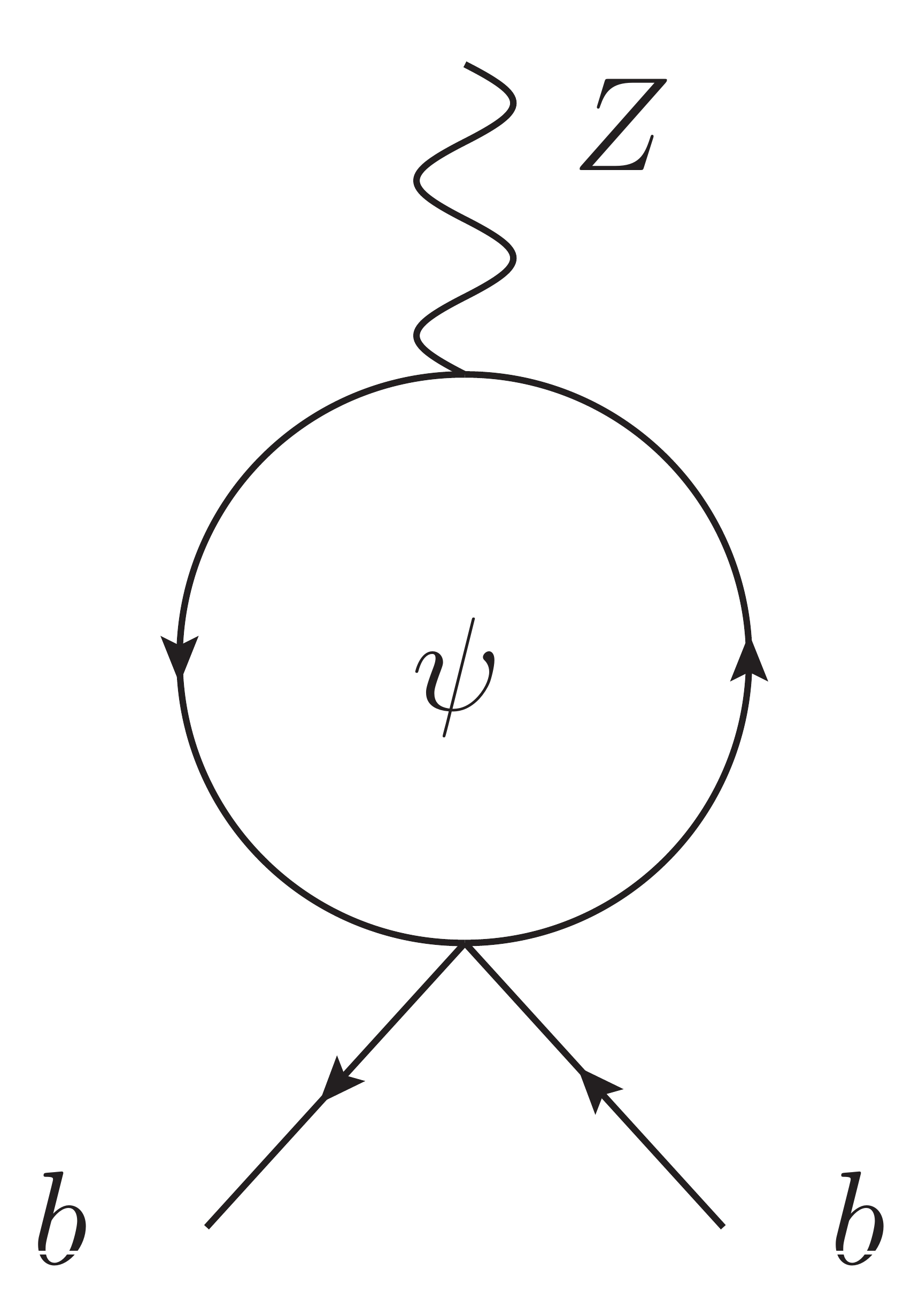}
\caption{Diagram contributing to the $Zbb$ coupling at finite momentum.}
\label{fig:bubble}
\end{figure}

From the point of view of the EFT below the scale of composite vectors, the dominant corrections to the coefficients of the operators~(\ref{eq:lag_zbb_p2}) come from the diagram depicted on Fig.~\ref{fig:bubble} with one four-fermion vertex, containing a flavour- and color-disconnected loop.
We will consider the following four-fermion interactions
\be
{\cal L}_{4f} = -{1 \over f_{\pi}^2} \sum_{p1,p2} C_{\rho} J_{\rho \, \mu}^{(p_1)} J_{\rho}^{(p_2)\, \mu} 
-{1 \over f_{\pi}^2} \sum_{p1,p2} C_{X}^{(p_1,p_2)} J_{X \, \mu}^{(p_1)} J_{X}^{(p_2)\, \mu}
\label{eq:4f_ops_zbbp2}
\ee
where $p1,p2=\{t,b\}$ and $J_{\rho,X\, \mu}^{(p)}$
are respectively the $SO(4)$ and $U(1)_X$ currents of composite $p$ partners. The coefficients $C$ are expected to be of the order one, we will take  $C_X^{(p_1 p_2)}$ symmetric in $p_1,p_2$. If the operators of Eq.~(\ref{eq:4f_ops_zbbp2}) were obtained by integrating out one layer of $SO(4)$ and $U(1)_X$ resonances, one would obtain  $C_{\rho}=1$ and $C^{(p_1,p_2)}_{X}=q_X^{p_1} q_X^{p_2}$ where $q_X=2/3,-1/3$ for the top and bottom partners respectively. This set of values will be called a benchmark set in the following.

\begin{figure}[t]
\centering
\includegraphics[width=0.45\textwidth]{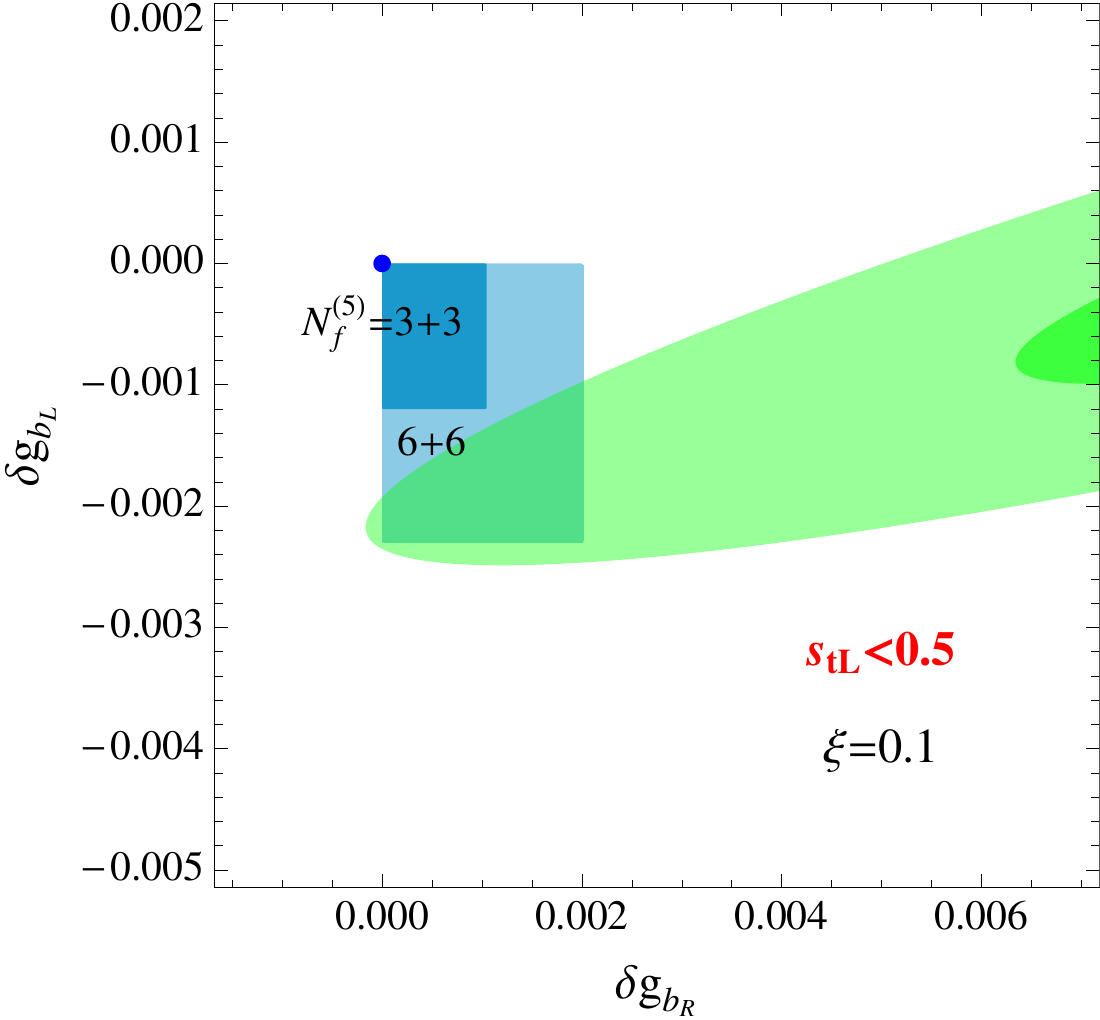}
\hfill
\includegraphics[width=0.45\textwidth]{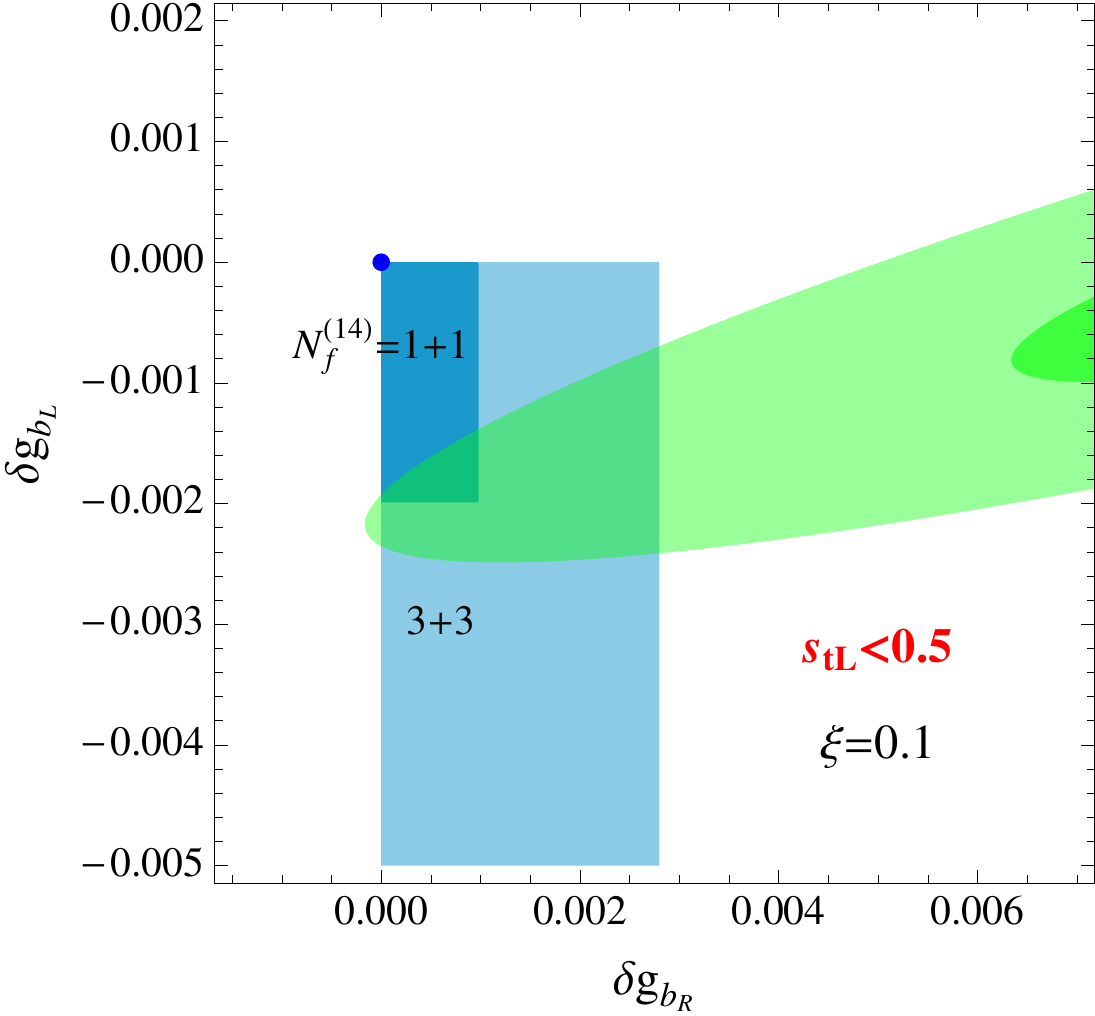}
\caption{Corrections to $Z\bar bb$ couplings induced at $q^2\ne0$ (in blue), for a benchmark set of four-fermion operators, for $\xi=0.1$ and the cutoff scale $3$~TeV, with the fermions (up+down partners) belonging to $[\bf 4+\bf 1]_{2/3}+[\bf 4+\bf1]_{-1/3}$ (left panel) and $[\bf 9+\bf 4+\bf 1]_{2/3}+[\bf 9+\bf 4+\bf 1]_{-1/3}$ (right panel) representations, with mass $m_{\psi}=1$~TeV, for different number of partners families (up + down) as indicated on plots. The mixing $s_L$ is limited to be at most $0.5$. Green area is experimentally allowed at $1\sigma$ (darker) and $2\sigma$ (lighter).}
\label{fig:Z_bb_p2_vect}
\end{figure} 

With the four-fermion operators~(\ref{eq:4f_ops_zbbp2}), assuming the equal number $N_f$ of the top and bottom partners families, we obtain the corrections to the $Z$ couplings
\bea
\delta g_{b_L}^{\text{($q^2$,1loop)}} &=&
s_{tL}^2 \,    N_c N_f \,    \left\{ -{1\over 2}(1-2 \sw^2) (2 C_{\bf r}) C_{\rho} -{\sw^2} d_{\bf r} \left({2\over3} C^{(t,t)}_{X}-{1\over3} C^{(b,t)}_{X}\right)\right\} 
{m_{Z}^2 \over 6 \pi^2 f_{\pi}^2} \log\left(m_{\rho,X}^2 \over m_{\psi}^2 \right) \nn\\
\delta g_{b_R}^{\text{($q^2$,1loop)}} &=&
 s_{bR}^2 \, N_c N_f  \,   \left\{ - \sw^2 d_{\bf r} \left( -{1\over 3} C^{(b,b)}_X+{2\over 3} C^{(b,t)}_X\right) \right\}
{m_{Z}^2 \over 6 \pi^2 f_{\pi}^2} \log\left(m_{X}^2 \over m_{\psi}^2 \right)
\eea
where $d_{\bf r}$ is a dimensionality of composite fermion representation, $C_{\bf r}$ is its Casimir operator with respect to $SO(4)$.
We can make a numerical estimate of these corrections assuming that the fermions have a common mass $m_{\psi}=1.5$~TeV and vectors $m_{\rho,X}=3$~TeV, taking the benchmark set of coefficients $C_{\rho,X}$
\bea
\delta g_{b_L}^{\text{($q^2$,1loop)}} 
& \simeq & -\{1., \, 5. \} \cdot 10^{-3} \, s_{tL}^2 \, N_f  \\
\delta g_{b_R}^{\text{($q^2$,1loop)}}
& \simeq & \{0.2, \, 0.6 \} \cdot 10^{-3} \, s_{bR}^2 \, N_f
\eea
where the values in brackets are given for the {\bf 1+4} and {\bf 1+4+9} composite fermions multiplets respectively. 
Both corrections are correlated in sign with SM couplings.
The correction to the right-handed coupling is suppressed by the $g_{b_R}^{\text{SM}}$, while $\delta g_{b_L}$ can be relatively large. On Fig.~\ref{fig:Z_bb_p2_vect} we present the estimated size of momentum-dependent corrections for the \FLFR and \FLOR models depending on the number of composite families below the vector resonances mass $m_{\rho}=3$~TeV, assuming for them common mass $m_{\psi}=1$~TeV, for $\xi=0.1$ and the coefficients of the four-fermion operators corresponding to the benchmark set. For this choice of parameters 3 families of up partners and 3 families of down-type partners in \FLFR case, as well as just one family of up and one of down-partners in case of \FLOR can generate a sufficient shift of $Z$ couplings in order to satisfy both $\sim 0$ deviation in low-energy measurements and $\sim - {\text{few}} \cdot 10^{-3}$ in $Z$-pole couplings. 

\subsubsection*{Corrections at zero momentum transfer}

Another type of corrections can be induced at zero momentum at one loop level, due to the mixing of charge 2/3 states with $t_L$ which violates $P_{LR}$ symmetry. 
We report here the analytic expressions for the loop-induced corrections separately for each composite multiplet, in the limit $g, g^{\prime}\to 0$, i.e. only considering the corrections due to the strong dynamics. In all the cases except the singlet, the four-fermion operators can induce corrections to $Z \bar bb$ couplings, which we will not report since they carry an additional model-dependence and have a slightly complicated form.

\subsubsection*{5+5}

\begin{itemize}

\item
{\bf singlet}
\begin{equation}
\delta g^{5+5(1)}_{b_L} = \frac{\xi}{64 \pi^2} \frac{y_{L1}^4 m_1^2 f_{\pi}^2}{(m_1^2 + y_{R1}^2 f_{\pi}^2)^3}
\left\{m_1^2 + 2 y_{R1}^2 f_{\pi}^2 \left[\log\left(\frac{2 (m_1^2 + y_{R1}^2 f_{\pi}^2)^2}{v^2 y_{L1}^2 y_{R1}^2 f_{\pi}^2}\right) - 1\right]\right\}
\label{eq:gbL55_1}
\end{equation}

\item
{\bf fourplet}
\begin{eqnarray}
\delta g^{5+5(4)}_{b_L} &=& -\frac{\xi}{32 \pi^2} \frac{y_{L4}^2 y_{R4}^2 f_{\pi}^2}{m_4^2 + y_{L4}^2 f^2}\Bigg\{\frac{y_{L4}^2 f_{\pi}^2}{m_4^2 + y_{L4}^2 f_{\pi}^2} + \left(1 - \frac{y_{R4}^2 f_{\pi}^2}{4 m_4^2}\right)
\log\left(1 + \frac{y_{L4}^2 f_{\pi}^2}{m_4^2}\right)\nonumber\\
&& - y_{L4}^2 f_{\pi}^2\frac{4 m_4^2(m_4^2 + y_{L4}^2 f_{\pi}^2) - (2 m_4^2 + y_{L4}^2 f_{\pi}^2) y_{R4}^2 f_{\pi}^2}{4 m_4^2 (m_4^2 + y_{L4}^2 f_{\pi}^2)^2}
\log\left(\frac{2 (m_4^2 + y_{L4}^2 f_{\pi}^2)^2}{v^2 y_{L4}^2 y_{R4}^2 f_{\pi}^2}\right)\Bigg\} \;\;\;\;\;\; \label{eq:Zbb4plet}
\label{eq:gbL55_4}
\end{eqnarray}

\end{itemize}

\subsubsection*{14+1}

\begin{itemize}

\item
{\bf singlet} is expected to give a positive correction  
\be
\delta g_{b_L}^{14+1 (1)} =  {\xi \over  64 \pi^2} {y_{L1}^2 f_{\pi}^2 \over m_1^2} \left\{ y_{L1}^2 + 2 y_{Lt}^2 \left[ \log\left({ m_1^2 \over m_{top}^2 }\right) -1 \right] \right\}
\label{eq:gbL141_1}
\ee

\item
{\bf fourplet}-induced correction can be both positive or negative
\bea
\delta g_{b_L}^{14+1 (4)} &=&-\frac{\xi}{64 \pi^2} \frac{y_{L4} f_{\pi}^2}{ (m_4^2 + y_{L4}^2 f_{\pi}^2)^2}\Bigg\{
m_4^2 y_{Lt}^2 (3 y_{L4} - \sqrt 2 c_{4t} y_{Lt} ) \nn \\
&& + {m_4^2 \over 2 f^2} (m_4^2 + y_{L4}^2 f_{\pi}^2) (5 y_{L4}-2 \sqrt 2 c_{4t} y_{Lt}) \log \left( {m_4^2 + y_{L4}^2 f_{\pi}^2 \over m_4^2 }\right) \nn \\
&& - m_4^2 y_{Lt}^2 \left[ y_{L4} \left( 2 + {y_{Lt}^2 f_{\pi}^2 \over 2(m_4^2 + y_{L4}^2 f_{\pi}^2)}\right) - \sqrt 2 c_{4t} y_{Lt} \right] \log \left( m_4^2 + y_{L4}^2 f_{\pi}^2 \over {m_{top}^2}\right)
\label{eq:gbL141_4}
\eea

\item
{\bf nineplet} can give both positive or negative correction depending on a size of $y_{L9}$. A small negative correction can be compatible with a small positive shift in $\hat T$ (see Eq.~(\ref{eq:T141_9})) 
\be
\delta g_{b_L}^{14+1 (9)}= {\xi \over  64 \pi^2} {y_{L9}^2 \over m_9^2} \left\{ 
y_{L9}^2 + 2 y_{Lt}^2 \left[3- \log \left( {m_9^2 \over m_{top}^2} \right) \right]
\right\}
\label{eq:gbL141_9}
\ee

\end{itemize}

\subsubsection*{Final remarks}


The four-fermion operators can in principle induce sizable corrections to momentum-independent shifts of $g_{b_L}$. The result however will strongly depend on the assumptions about the size and signs of four-fermion operators (see Ref.~\cite{ewpt}, also for other types of higher-dimensional operators potentially important for $g_{b_L}$). Moreover, some of the contributions induced by four-fermion operators are logarithmically sensitive to the cutoff~\cite{ewpt}. Given this, we will prefer not to use them in our numerical analysis. Nevertheless we will use the other loop-order corrections just to estimate their typical size and to understand which kind of additional contributions is needed (if needed) for the considered scenarios to pass the experimental constraints. 
The latter are quite ambiguous and one can take different attitude towards them, admitting the apparent deviation or assume it to be an artifact. In the following numerical scan we will require $|\delta g_{b_L}|<10^{-3}$, just to indicate the areas where the  
corrections are minimal.

Given our decision not to account for four-fermion operators in $\delta g_{b_L}$ at zero momentum, it would not be consistent to account for them in case of the momentum-dependent corrections. The latter ones are even more ambiguous, being sensitive even to the composite states not coupled to the quarks. Nevertheless it is still important that in principle CH models can explain the deviation of the $Z\to \bar b b$ decay parameters without distorting lower energy $Z$ couplings, like for instance those relevant for meson decays which we discuss in the next Section.  

\subsection{$Z$-mediated FCNC in down sector}

The important constraints on the flavour-changing $Z$-boson couplings in CH models come from the flavour physics observables~\cite{Straub:2013zca}.
The bounds imposed by the $Z$-pole measurements instead are quite weak (see for instance Ref.~\cite{Atwood:2002ke}). 
Hence in this Section we will only analyse the corrections to the $Z \bar d^i d^j$ vertices at zero momentum transfer.
In this case the up sector, which we focus on in this work, can only generate the loop-level corrections, which we parametrize as 
\be
{\cal L}^Z_{\text{FCNC}} = \frac{g}{\cw} Z_\mu \overline d^i \gamma^\mu \left[
\delta g_{L}^{i j} P_L + \delta g_{R}^{i j} P_R\right] d^j\,.
\ee 
One of the most stringent constraints on $Z$-mediated FCNC comes from the measurement of the  $B_s \to \mu^+ \mu^-$ decay branching ratios.
The analysis presented in Ref.~\cite{Guadagnoli:2013mru} has shown that the constraints imposed on flavour-changing $Z$ couplings, 
under certain assumptions on the flavour structure (anarchic partial compositeness or generic MFV), can bound flavour-diagonal $Z\bar bb$ coupling in a comparable way with the LEP $Z$-pole measurements.
The recent improvements in the constraints on flavour-averaged branching ratios of $B_s$-meson decays~\cite{Buras:2013qja,bsmumuexp}
\be
\overline{\cal B}^{\text{exp}} = (2.9 \pm 0.70) \cdot 10^{-9} \, , \;\;\;\;  \overline{\cal B}^{\text{SM}} = (3.56 \pm 0.18) \cdot 10^{-9} 
\label{eq:bsmumubounds}
\ee
further squeeze corresponding bounds. In order to translate these results into the bound on $\delta g_{L,R}^{i j}$ we will use an expression for the modified $B_s \to \mu^+ \mu^-$ branching ratio~\cite{Guadagnoli:2013mru}
\be
{\cal B}(B_s \to \mu^+ \mu^-)={\cal B}^{\text{SM}}(B_s \to \mu^+ \mu^-) \left| 1+{\sqrt 2 \pi^2 \over G_F m_{\text{W}}^2V_{tb}^{\star}V_{ts}} {(\delta g_L^{32}-\delta g_R^{32})\over Y_{\text{SM}}} \right|^2 \,,
\label{eq:bs_br}
\ee
with $Y_{\text{SM}}\simeq 0.957$. The expression for ${\cal B}(\bar B_s \to \mu^+ \mu^-)$, needed to compute the flavour-averaged BR of Eq.~(\ref{eq:bsmumubounds}), can be obtained after an exchange $b \leftrightarrow s$ in Eq.~(\ref{eq:bs_br}). 

In the following we will derive the couplings of the $Z$-mediated FCNC, expressing them in terms of flavour-conserving $Z$ couplings, for each of the considered flavour pattern.  We will only consider the left-handed couplings since the right-handed ones do not receive any corrections from the up sector-related composite dynamics.


\begin{itemize}

\item
$\bf{U(3)^2}$ {\bf LC}: in this case the flavour breaking is induced by $y_R^{i j}$ (explicitly) and by $d_L^i$ (implicitly, originating from the source of down-type Yukawas). The mixings in the mass eigenstate basis can be brought to a form
\bea
\bar q_L y_{L} &\to & 
\begin{cases}
\bar{\hat u}_L  y_{L}  & (a)\\
\bar{\hat d}_L V_{\text{CKM}}^{\dagger} y_{L} & (b)
\end{cases} \label{zfcnc_U3L_1}\\
\bar u_R y_R &\to& \bar{\hat{u}}_R \hat y_R \sim \{0,0,\bar{\hat{t}}_{R} \hat y_{tR} \}
\label{zfcnc_U3L_2}
\eea
Since $y_L$ is proportional to identity, the CKM matrix from the mixing~(\ref{zfcnc_U3L_1}, (b)), when used to compute the $Zd_id_j$ coupling, can be removed and contracted with the $V_{\text{CKM}}$ originating from the opposite fermionic leg, unless the contribution contains $\hat y_{R}$. In this latter case  the FCNC proportional to $\xi^{i j}$ are generated. Therefore flavour-changing couplings of the $Z$ can be expressed as
\be
\delta g_{L}^{ij} = \xi^{i j} \left(\delta g_{b_L}-\delta g_{d_L} \right)
\ee 
given that the difference between $g_{b_L}$ and $g_{d_L}$ is only caused by a sizeable $\hat y_{tR}$. 

\item
$\bf{U(3)^2}$ {\bf RC}: breaking resides in $y_L^{i j}$ (explicitly) and in $d_L^i$ (implicitly). Using $U(3)$ rotations in the composite and $d_R$ sectors we can bring the left mixings to the form
\be
\bar q_L y_{L} \to  
\begin{cases}
\bar u_L U^{\dagger}_{u} \hat y_{L} \to \bar{\hat u}_L\hat y_{L} & \sim \{0,0, \bar{\hat t}_{L} \hat y_{tL} \}  \\
\bar d_L U^{\dagger}_{u} \hat y_{L}  \to \bar{\hat d}_L V_{\text{CKM}}^{\dagger} \hat y_{L} &\sim \{0,0, \bar{\hat d}_{iL} V_{\text{CKM} \, i 3}^{\dagger} \hat y_{tL} \}
\end{cases}
\ee 
while the right mixings are proportional to identity. It is thus straightforward to derive the relation between leading flavour-changing and flavour-conserving modifications of the $Z$ couplings:
\be
\delta g_{L}^{ij} = \xi^{i j} \delta g_{b_L}
\ee 


\item
$\bf{U(2)^2}$ {\bf LC}, $t_R$ partially composite: invoking the explicit form of the $t_L$ and $b_L$ rotations in flavour space (see Ref.~\cite{Barbieri:2011ci}) we obtain
\be
\delta g_{L}^{ij} \simeq \xi^{i j} \kappa^{i j} \delta g_{b_L}\, , \;\;\; \kappa^{i3}=r_b, \kappa^{12}=|r_b|^2
\label{eq:dgij_u2lc}
\ee 
where $r_b$ is a complex parameter defined in Ref.~\cite{Barbieri:2012tu}. The coefficient $r_b$ depends on the details of the down-quark mass generation, and its absolute value can be made smaller than 1, making the bound from $B_s\to \mu^+ \mu^-$ comparable or less restrictive than the one from $Z\to \bar bb$. 
Notice, that $r_b$ can also suppress tree-level FCNC induced by four-fermion operators which will be discussed later on.  

\item
$\bf{U(2)^2}$ {\bf RC}, $t_R$ partially composite: again using the explicit form of rotation matrices we obtain
\be
\delta g_{L}^{ij} = \xi^{i j} \delta g_{b_L}
\label{eq:dgij_u2rc}
\ee 

\item
$\bf{U(2)^2}$ {\bf TC} with a partially composite $t_R$ and $\bf{U(2)^2}$ with a totally composite $t_R$ lead to the relation given by Eq.~(\ref{eq:dgij_u2lc}).

\end{itemize}

\begin{figure}
\centering
\includegraphics[width=.46\textwidth]{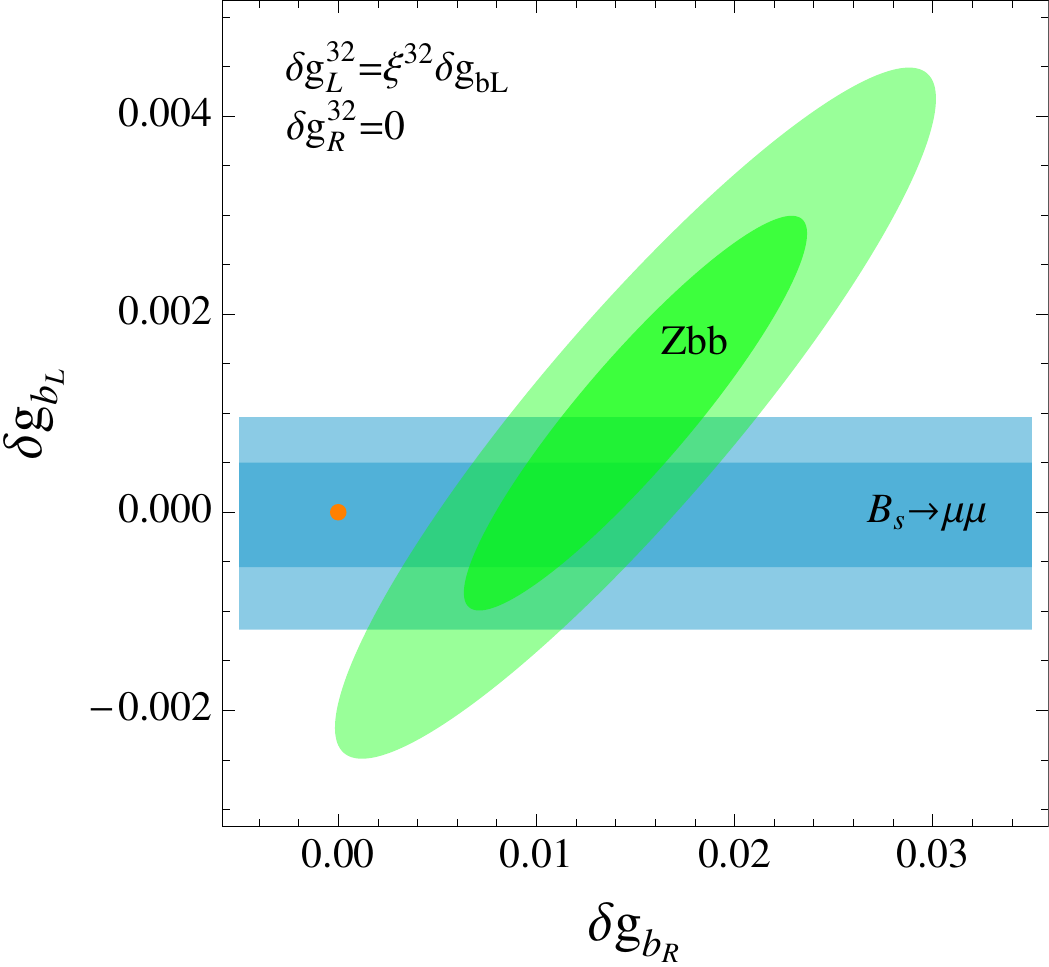}
\caption{Constraints on the corrections to the $Z$ boson couplings to the bottom quark, imposed by $Z \to \bar bb$ (green ellipses) and $B_s \to \mu^+ \mu^-$ (blue bands),
at $68\%$CL (darker) and $95\%$CL (lighter), assuming $\delta g_L^{32}=\xi^{32}\delta g_{b_L}$. Orange point corresponds to the SM prediction. Notice, however, that $\delta g_{b_L}$ can have different meanings in the two cases, see text for details.}
\label{fig:deltag_bsmumu}
\end{figure}

On Fig.~\ref{fig:deltag_bsmumu} we present the constraints on $\delta g_{b_L}$ and $\delta g_{b_R}$, derived from the $A_{\text {FB}}^b$ and $R_b$ measurements (in green) and deduced from the $B_s\to \mu \mu$ branching (in blue) assuming the relations $\delta g_{L}^{i j} = \xi^{ij}\delta g_{b_L}$ and $\delta g_{R}^{i j}\simeq0$, which is the case for $U(3)^2_{\text{RC}}$, $U(2)^2_{\text{RC}}$ and also for the other $U(2)^2$ patterns if $r_b=1$.  Notice that even if sizeable $\delta g_{b_R}$ is not generated in the model, the two constraints can still be in agreement if a sizeable negative correction to $\delta g_{b_L}$ is produced by the operator~(\ref{eq:lag_zbb_p2}), which is relevant for $Z$ decays but not for $B_s$.

For what concerns the numerical analysis of this paper, for our choice of the bound $|\delta g_{b_L}| < 10^{-3}$, the limit arising from $B_s\to \mu \mu$ coincides with the one from $Z \to b b$ for RC, and can be less or more constraining for other $U(2)^2$ scenarios, depending on the value of $r_b$. In the following we will use $r_b=1/2$, hence this constraint will be irrelevant.   

The relations obtained in this Section can in principle be violated in a presence of sizeable four-fermion interactions, containing quarks and leptons, which can contribute to $B_s \to \mu\mu$ without modifying $Z$ couplings. However it is reasonable to assume that this type of operators is highly suppressed by small lepton masses.    

\paragraph{}

We will also briefly comment here on two other types of FCNC.  
In all the considered scenarios Higgs-mediated FCNC do not arise in the order $\sim y_L y_R$ (see Ref.~\cite{Azatov:2014lha} for more details). 
The leading contributions to FCNC appear in $y^4$ order and are naturally suppressed by the small masses of light quarks. Driven by this argument we neglect the impact of possible operators of this type. 
Another type of FCNC, the loop-induced chirality-breaking effects, do not bring relevant constraints on the considered CH scenarios with flavour symmetries, given one imposes $CP$ conservation in the composite sector (see Ref.s~\cite{Barbieri:2012tu,Konig:2014iqa}).

\subsection{Tree-level $\Delta F=2$ operators}

\subsubsection*{Partial Compositeness}

In all the scenarios considered in the present paper one of the dominant contributions to the tree-level $\Delta F =2$ processes comes from the four-composite-fermion operators discussed in Sec.~\ref{sec:EFT_five}. 
After accounting for elementary-composite mixings they give rise to flavour-changing four-quark interactions. The most constraining FCNC processes are those containing down-type quarks, which have sizeable mixings only with $B_L$ resonances.
The fermionic currents, entering the four-fermion operators, can be transformed as 
\be
\bar B_L \gamma_{\mu} B_L \to \bar b_L s_L \gamma_{\mu} s_L^{\dagger} b_L \to \bar{\hat{b}}_L U_{bL}^{\dagger} s_L \gamma_{\mu} s_L^{\dagger} U_{bL} \hat b_L.
\ee
Using the results for bilinears operators obtained in the previous Section, we can write down the $\Delta F=2$ operators as  
\be
{1 \over f_{\pi}^2} \, \xi_{i j}^2 \, \kappa_{i j}^2 \, s_{tL}^{4} C_{dL} [\bar d_i \gamma_{\mu} d_j] [\bar d_i \gamma^{\mu} d_j].
\label{eq:4ferm_3} 
\ee
The coefficient $\kappa$ depends on the flavour pattern:
in $U(3)^2_{\text{LC}}$ $\kappa_{i j}=0$; in $U(3)^2_{\text{RC}}$ and $U(2)^2_{\text{RC}}$  $\kappa_{i j}=1$; in $U(2)^2_{\text{LC}}$, $U(2)^2_{\text{TC}}$ and $U(2)^2_{t_R\text{comp}}$ $\kappa^{i3}=r_b$ and $\kappa^{12}=|r_b|^2$.
The coefficient $C_{dL}$ is expected to be of the order one. The resulting bound on  $s_L$ depends on $(C_{dL})^{1/4}$, hence it is rather insensitive to the uncertainty on the coefficient size. 
For the bounds on the operator~(\ref{eq:4ferm_3}) we take those obtained in Ref.~\cite{Calibbi:2012at}. Choosing the sign of $C_{dL}$ to maximize the constraints, we obtain
\be
U(3)^2_{RC}, U(2)^2_{RC}: s_L \lesssim 0.39_{(1,2),(2,3)} , 0.42_{(1,3)} \, ,
\label{eq:bound_s_L_fcnc}
\ee
where the subscripts $(i,j)$ stand for flavour indices of the four-fermion operator for which the constraint is obtained.
In $U(3)^2$ LC this type of bound is absent, while in the other scenarios the exact constraint depends on $r_b$.

\subsubsection*{Technicolor}

Additional four-quark operators can be generated in $U(2)^2$ TC scenario at the scale $\Lambda^{\prime}$, together with the operators responsible for light quark masses. At low energies the relevant lagrangians are
\be
{\cal L}_{\text {TC}}^{\text{FCNC}} \sim {1\over \Lambda^{\prime \,2}} q_L^4 \, , \;\;\;\;\;\;\;\;\;\; 
{\cal L}_{\text {TC}}^{\text{mass}} = \lambda_{1,2} \,  \bar q^{\bf5}_L {\cal O}  u_R^{\bf5} \to \lambda_{1,2}  {v \over \sqrt 2} \bar q_L u_R \, , 
\label{eq:lag_tc_fcnc}
\ee
The strongest lower bound $\Lambda^{\prime} \gtrsim 10^5$~TeV can be obtained from the constraints on $CP$ violating processes, assuming that the FCNC Lagrangian~(\ref{eq:lag_tc_fcnc}) generates them with order-one strength. 

In principle the same type of four-quark operators can also be expected in the scenarios with a partial compositeness only, but in that case it is known that the scale $\Lambda^{\prime}$ can be made arbitrarily large without conflicting with the quark masses~\cite{Contino:2010rs}. 
In the TC case, instead, the high UV scale $\Lambda^{\prime}$ generically also suppresses the TC-like masses of the quarks. 
Indeed, from the form of the mass Lagrangian of Eq.~(\ref{eq:lag_tc_fcnc}) follows that the strength of the Yukawa interactions $\lambda_{1,2}$ at the strong dynamics confinement scale $\Lambda \lesssim 4 \pi f_{\pi}$ should be 
\be
\lambda_{1,2} \lesssim  4 \pi  \left( \Lambda \over \Lambda^{\prime} \right)^{[{\cal O}]-1}\,,
\label{eq:yuk_tc}
\ee
where the operator ${\cal O}$ is assumed to have a scaling dimension $\sim[{\cal O}]$ over the interval of energies from $\Lambda^{\prime}$ to $\Lambda$. This constraint is dictated by a requirement of $\lambda_{1,2}$ perturbativity at $\Lambda^{\prime}$. The scaling dimension of the composite scalar operator $[{\cal O}]$ is generically expected to be greater than 1, hence one can expect difficulties with reproducing the top mass. Moreover, $[{\cal O}]$ very close to 1 is incompatible with having a scaling dimension of the $SO(5)$-singlet operator ${\cal O}^2$ close to or larger than 4. Presence of the relevant ${\cal O}^2$ operator would make the scale $\Lambda$ too sensitive to the scale $\Lambda^{\prime}$, making a large separation between them unnatural.  In order to keep $\Lambda$ stable one needs $[{\cal O}] \gtrsim 1.5$~\cite{Rattazzi:2008pe,Rychkov:2009ij, Rattazzi:2010yc,Vichi:2011ux,Poland:2011ey}~\footnote{We thank Slava Rychkov for drawing our attention to the latest works on this subject.}.

The advantage of the mixed TC scenario is that the largest mass generated by the TC-like interactions is the mass of the charm. As follows from Eq.~(\ref{eq:yuk_tc}), the charm mass allows for $\left(\Lambda \over \Lambda^{\prime} \right) \gtrsim 10^{-6}$, which for $\Lambda \sim 10$~TeV leads to $\Lambda^{\prime} \lesssim 10^7$~TeV.  
Hence all the FCNC generated at $\Lambda^{\prime}$ can be safely ignored.

\subsection{Compositeness constraints}

In the flavour patterns, where the degree of light quark compositeness is related to the one of the top quark, stringent constraints come from the searches for quark compositeness. 
The four-fermion interactions in the composite sector can generate the following four-quark operators
\be
{s_L^4 \over f_{\pi}^2}    [\bar q_L \gamma_{\mu} q_L] [\bar q_L \gamma^{\mu} q_L] \,, \;\;\; 
{s_L^4 \over f_{\pi}^2}    [\bar q_L \gamma_{\mu} T^A q_L] [\bar q_L \gamma^{\mu} T^A q_L] \,, \;\;\;
{s_R^4 \over f_{\pi}^2}    [\bar u_R \gamma_{\mu} u_R] [\bar u_R \gamma^{\mu} u_R] \,,
\label{eq:ops_dj}
\ee
where $T^A$ are $SU(3)_c$. We have listed the most constraining operators involving up and down quark in LC (first and second operators) and RC (third operator) scenarios~\cite{Barbieri:2012tu}.
Such type of interactions leads to deviations in the angular distributions of jets with respect to the SM predictions, and thus one can put the following bounds on the corresponding operators coefficients~\cite{Domenech:2012ai}, for the case when $q_L$ and $u_R$ are the first generation quarks
\be
{1\over (5.0\,\text{TeV})^2} \;\; , \;\;\;\; {1\over(3.4\,\text{TeV})^2} \;\; , \;\;\;\; {1 \over(4.5 \,\text{TeV})^2} \, ,
\ee
which for $\xi=0.1$ translates into the bounds on the mixing angles (assuming that the coefficients of operators~(\ref{eq:ops_dj}) are equal to one)
\be
s_L , s_R \lesssim 0.4.
\ee
The bounds on $s_L$ and $s_R$ apply to the scenarios with $U(3)^2$ with the  Left and the Right Compositeness respectively. 

\subsection{Direct searches}

Currently the most stringent experimental constraints on fourplet and singlet top partners from the direct searches were derived in Ref.s~\cite{cms53_new,Chatrchyan:2013uxa}, they set limits of $800$~GeV and $\sim 700$~GeV on the masses of the charge 5/3 fourplet partner and the charge 2/3 singlet respectively. These bounds are model independent since they rely on the pair production mechanism mostly defined by QCD interactions. There already exist experimental searches for the singly produced partners~\cite{last} as well as recasts of the searches for pair production into the bounds on singly produced partners~\cite{DeSimone:2012fs,Azatov:2013hya,Matsedonskyi:2014mna,MPW2}, but for the current experimental sensitivity they do not lead to a significant improvement of the bounds, therefore we will not use them in the following.

Despite the absence of the dedicated searches for the states present in the nineplet, one can recast the results of analyses dedicated to charge 5/3 states into the bound on $m_9$. This bound, which is mainly driven by the charge 8/3 state, gives $m_9 \gtrsim 1$~TeV~\cite{Matsedonskyi:2014lla}.  

\begin{figure}
\centering
\includegraphics[width=.4\textwidth]{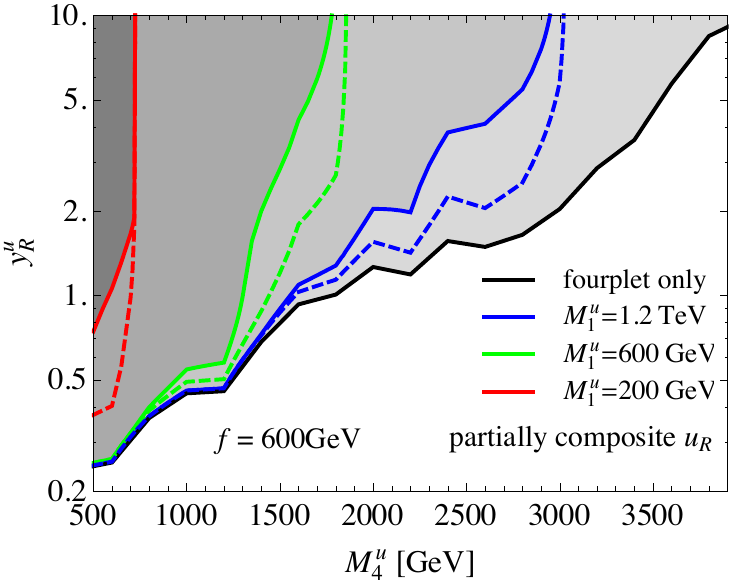}
\caption{95\% CL bounds on the fourplet mass parameter $m_4$ depending on the mixing with the right-handed up quark for $f=600$~GeV and $c_{41}=0$ and different choices of the singlet mass parameter~\cite{Delaunay:2013pwa}. 
Dashed lines account for the effect of reduction of production cross section of the fourplet partners in a presence of the singlet, while 
solid lines also include an effect of reduced branching fraction due to cascade decays.}
\label{fig:excl_yR_m4_up}
\end{figure}

Besides the top partners searches, the searches for the partners of light generations can also be important if partner mixings with light quarks are sizeable, which is the case for $U(3)^2$ symmetric scenarios. 
A study of the constraints arising in case of large degree of compositeness of right-handed up-type quarks was performed in Ref.s~\cite{Delaunay:2013pwa,Redi:2013eaa}.   
Fig.~\ref{fig:excl_yR_m4_up} taken from Ref.~\cite{Delaunay:2013pwa} shows a bound on the fourplet mass parameter $m_4$ depending on the mixing parameter $y_{R4}$ of the up-quark for $f_{\pi}=600$~GeV and four choices of a parameter $m_1$. Given that the leading couplings of the partners with gauge bosons and SM quarks scale as $v \over f_{\pi}$, one can easily obtain the bound corresponding to different values of $f_{\pi}$. This bound can be interpreted as conservative since it assumes $c_{41}=0$ which reduces the expected production cross section. The analogous bound in case of charm partners is much weaker and does not improve significantly the overall bound on $m_4$ if combined with the one on the up partners.     

The experimental searches for uncolored vectorial composite resonances (see for instance Ref.s~\cite{vectorsdirectsearches}) lead to a bound on their mass of around $2$~TeV for $\xi=0.1$~\cite{thamm2013,Pappadopulo:2014qza,Greco:2014aza}.


\section{Numerical results for $U(3)^2$}
\label{num3}

In this Section we present the results of the numerical analysis of the \FLFR model with $U(3)^2$ flavour symmetry, using the constraints discussed in the previous Section. 
First, we will summarize the most constraining observables for LC and RC flavour patterns:

\begin{itemize}

\item
In both cases the EWPT constraints are important, in the LC case the $\hat T$ parameter receives contributions from all three generations of partners, while for RC only the top partners are relevant. In both cases $\hat S$ is sensitive to all the partner generations due to loop effects.

\item
The bounds on $Z \bar b b$ couplings are in principle always relevant. But as we stated in Section~\ref{sec:Zbb}, we will just show the areas where the least model dependent part of the correction is small. 

\item
Tree-level $\Delta F=2$ FCNC are only relevant in RC case.

\item
Dijet constraints apply to both LC and RC cases, due to the fact that all three families of either left- or right-handed up-like quarks share the same degree of compositeness.    

\item
$\delta V_{ud}$ can only be sizable in LC case. 

\item
$R_h$ can receive tree-level corrections (from $Z \bar u u$ and $Z \bar c c$) and loop-level (from $Z \bar d d$, $Z \bar s s$ and $Z \bar b b$) in LC case, and only loop-level for RC (from $Z \bar b b$). Again, as in the case of $\delta g_{b_L}$ we will not consider the loop effects induced by four-fermion operators.

\end{itemize}

\begin{figure}
\centering
\includegraphics[width=.495\textwidth]{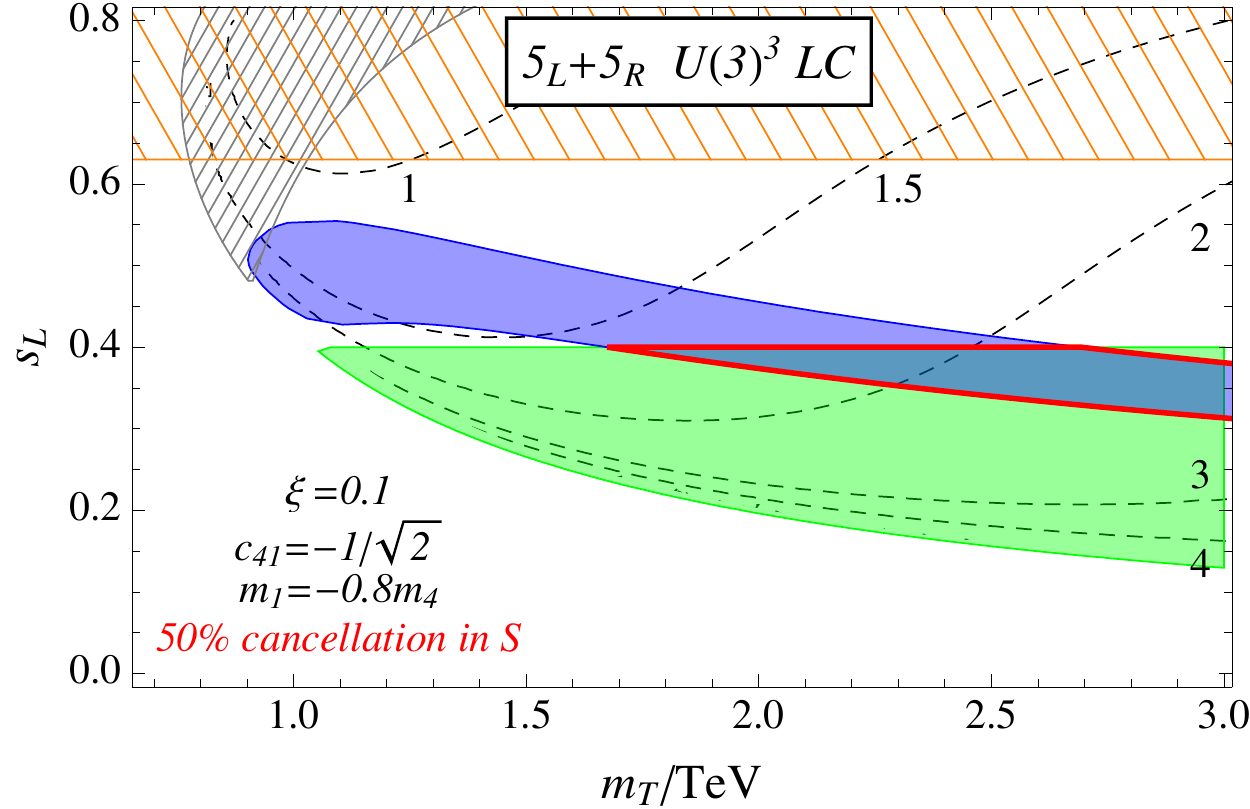}
\includegraphics[width=.495\textwidth]{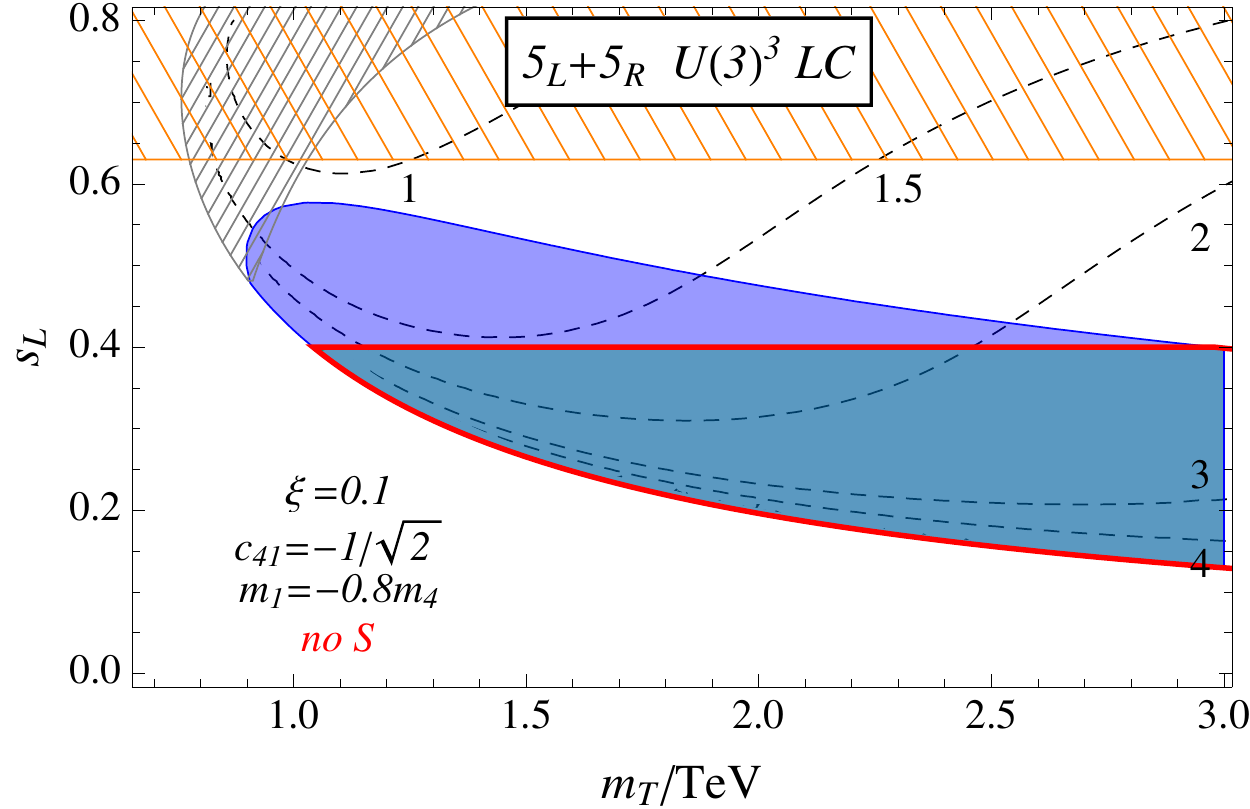}
\caption{Bounds on the \FLFR model with $U(3)^2$ LC flavour pattern, for $\xi=0.1$, $c_{41} = - 1/\sqrt 2$ and $m_1= - 0.8 m_4$, in $m_T-s_L$ plain. Dashed grey and orange areas are forbidden by direct searches and $\delta V_{ud}$ respectively, green area is allowed by dijets. On the left panel the blue are is allowed by $\hat S$ (with 50\% reduction) and $\hat T$, on the right panel the blue  area is allowed by $\hat T$ (marginalized over $\hat S$). The areas allowed by all the constraints are shown by red contours. Dashed lines are the singlet top partner ($\widetilde T$) mass isolines (mass shown in TeV).}
\label{fig:excl_U3_LC_5_5}
\end{figure}

Let us start with the LC scenario. In this case the stringent constraints from $\delta V_{ud}$ can only be avoided for $c_{41}\simeq - 1/\sqrt 2$ and $m_1 \simeq - m_4$ (see Sec.~\ref{sec:ZttWud}). We present the exclusion plots for this type of configuration on Fig.~\ref{fig:excl_U3_LC_5_5}, in $m_T-s_L$ plane, where $s_L=y_L f_{\pi} / m_T$. For these plots we set $c_{41} = - 1/\sqrt 2$, $m_1= - 0.8 m_4$, $\xi=0.1$ and $m_{\rho}=3$~TeV. The right plot shows the areas allowed by the most robust variables: $\hat T$ parameter (marginalized over $\hat S$), dijet constraints, $\delta V_{ud}$ and direct searches, showing that all the masses above $1$~TeV are allowed for the $T$ and above $1.5$~TeV for the $\widetilde T$. Adding 50\% of the contribution to $\Delta \hat S$, predicted by the considered model, pushes the minimally allowed partners masses to $1.7$~TeV. Therefore in order to keep the partners relatively light one needs additional compensating contributions to the $\hat S$ parameter, which can come for instance from the down partners or even from the leptonic ones. 
It is important to notice that this only region allowed by $\delta V_{ud}$ is characterised by $m_T\sim m_{\widetilde T}$, for which the Higgs mass can tolerate the heaviest top partners, up to $\sim 2$~TeV for $\xi=0.1$, therefore this parameter space area will be completely covered by the direct searches in the last turn. We did not show the $\delta g_{b_L}$ and $R_h$ constraints since in this case they are satisfied practically everywhere.

\begin{figure}
\centering
\includegraphics[width=.49\textwidth]{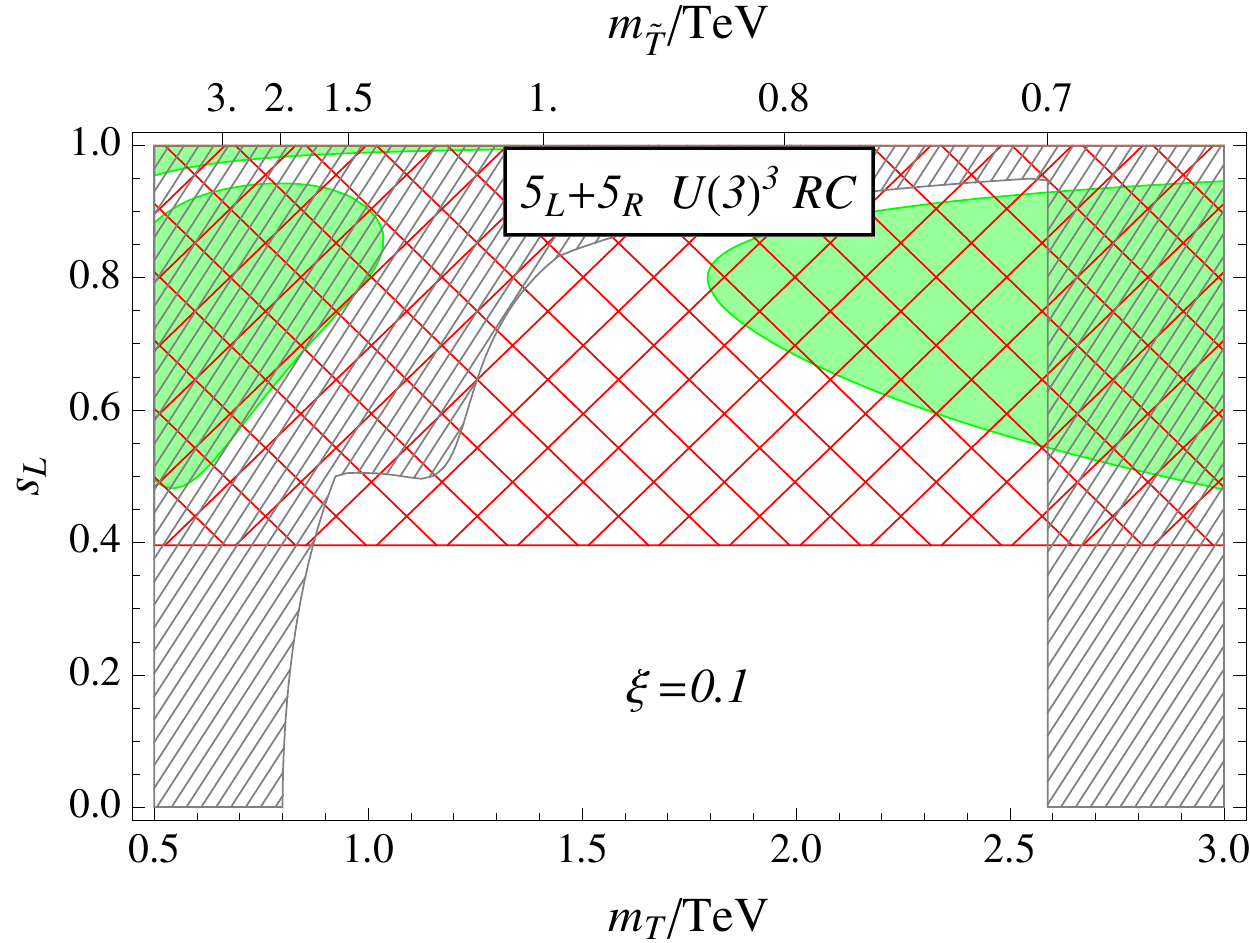}
\includegraphics[width=.499\textwidth]{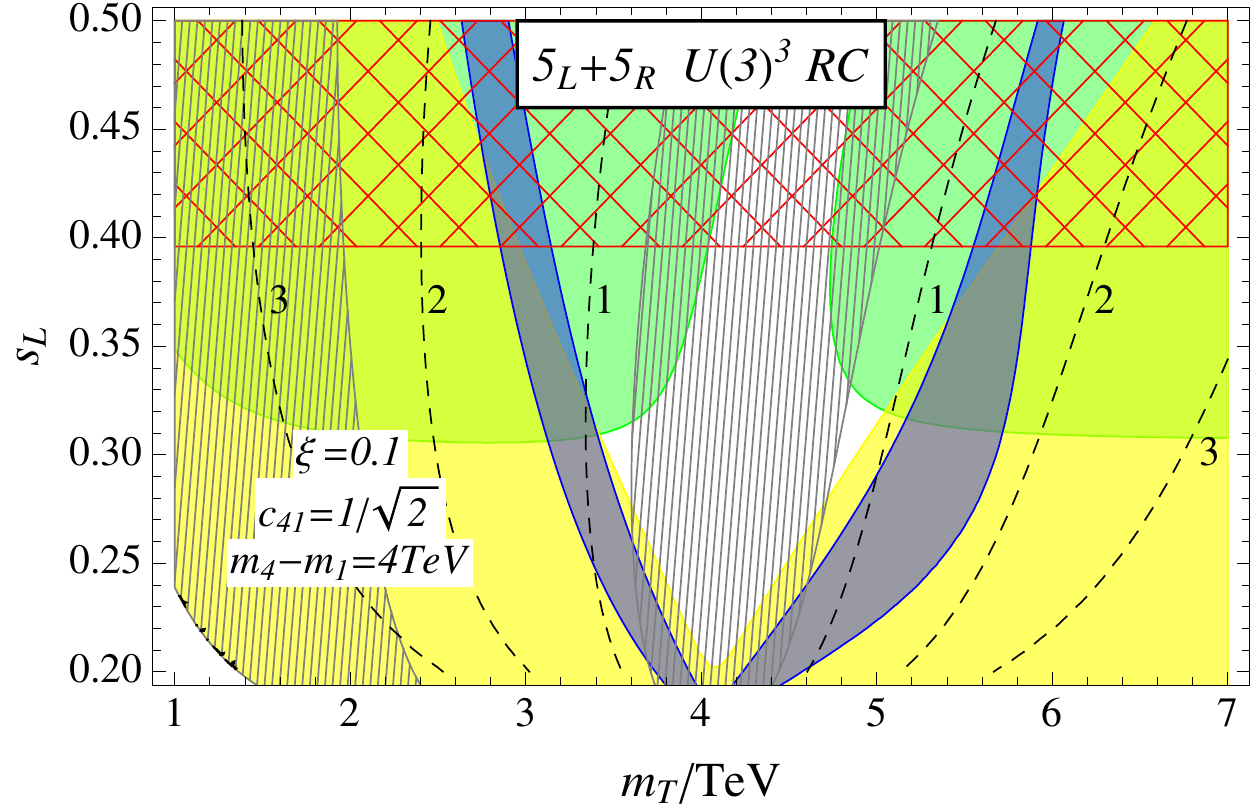}
\caption{Left panel: bounds on the \FLFR model with $U(3)^3$ RC flavour pattern, with the Higgs mass fixed by Eq.~(\ref{eq:mh_2s}), for $\xi=0.1$, axes correspond to the masses of the fourplet ($m_T$) and the singlet ($m_{\widetilde T}$) top partners and to the $t_L$ degree of compositeness $s_L$. Grey and red meshed areas are excluded by direct searches and $\Delta F=2$ respectively, green area is compatible with dijet bounds. 
Right panel: bounds on the same model, without fixing the Higgs mass, for $m_4-m_1=4$~TeV, $\xi=0.1$, $c_{41}=1/\sqrt2$ and a cutoff $m_{\rho}=10$~TeV. Grey and red meshed areas are excluded by direct searches and $\Delta F=2$ respectively. Blue area is allowed by $\hat S$ and $\hat T$,  green -- by dijets searches, yellow -- allowed by $\delta g_{b_L}$. Dashed black lines are $\widetilde T$ mass isolines (labels in TeV).}
\label{fig:excl_U3_RC_5_5}
\end{figure}

On Fig.~\ref{fig:excl_U3_RC_5_5} we present the most constraining bounds for the $U(3)^2$ RC scenario for $\xi=0.1$. On the left panel we fixed the Higgs mass by the condition~(\ref{eq:mh_2s}), imposed on the top partner masses. In this case all the range of partner masses is excluded by a combination of the dijet and $\Delta F=2$ constraints. 
These two constraints bound $s_{tL}$ and $s_{tR}$ such that one can not generate a large enough top Yukawa $\lambda_{top} \simeq s_L s_R {(m_4-m_1)/ f_{\pi}}$.
Abandoning the relation $m_h(m_T,m_{\widetilde T})$ of Eq.~(\ref{eq:mh_2s}), we can allow for larger $(m_4-m_1)$ hence generating $y_{top}$ even for small mixing angles.
On the right panel of Fig.~\ref{fig:excl_U3_RC_5_5} we present the bounds on the configuration with $(m_4-m_1)\sim 4$~TeV, $c_{41}=1/\sqrt 2$ and the maximal $m_{\rho}=10$~TeV~\footnote{The cutoff is set above all the fermionic mass scales since our analysis is only suitable for this configuration, but in this case the exact value of the cutoff does not play an important role.}. In this case a combination of $\hat S$, $\hat T$, dijets and FCNC bounds allows to have the light singlet $\widetilde T$ with a mass $1-2$~TeV, while the fourplet partners are always very heavy. This result is easy to understand given that the large difference $(m_4-m_1)$ requires either fourplets or singlets to be very heavy, but the light fourplets alone have difficulties with EWPT (negative shift in $\hat T$, large positive shift in $\hat S$).



\section{Numerical results for $U(2)^2$}
\label{num2}

We will now present the results of the combined analysis of the models with $U(2)^2$ flavour symmetry. 
The constraints on these scenarios are weaker than those on $U(3)^3$, in particular the dijet constraints are not relevant, as well as $R_h$ and $V_{ud}$. 
The $\hat T$ parameter is now only sensitive to the top partners, the $\hat S$ parameter and $\delta g_{b_L}$ can steel be affected by the light quark partners circulating inside the loops (but we will not analyse such effects in details).
Tree level $\Delta F=2$ FCNC arise in all the models, but only in the \FLFR RC case they can not be suppressed by the parameter $r_b$, which is sensitive to the details of the down-type sector.

\subsection*{\FLFR}

\begin{figure}
\centering
\includegraphics[width=.32\textwidth]{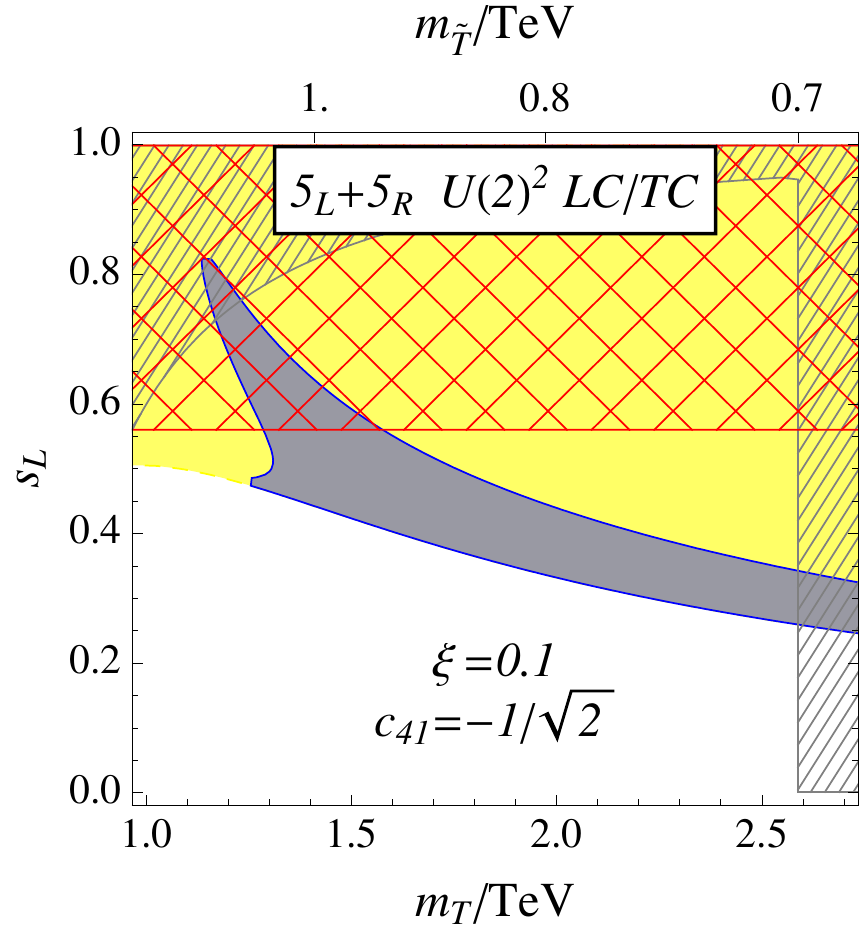}
\includegraphics[width=.32\textwidth]{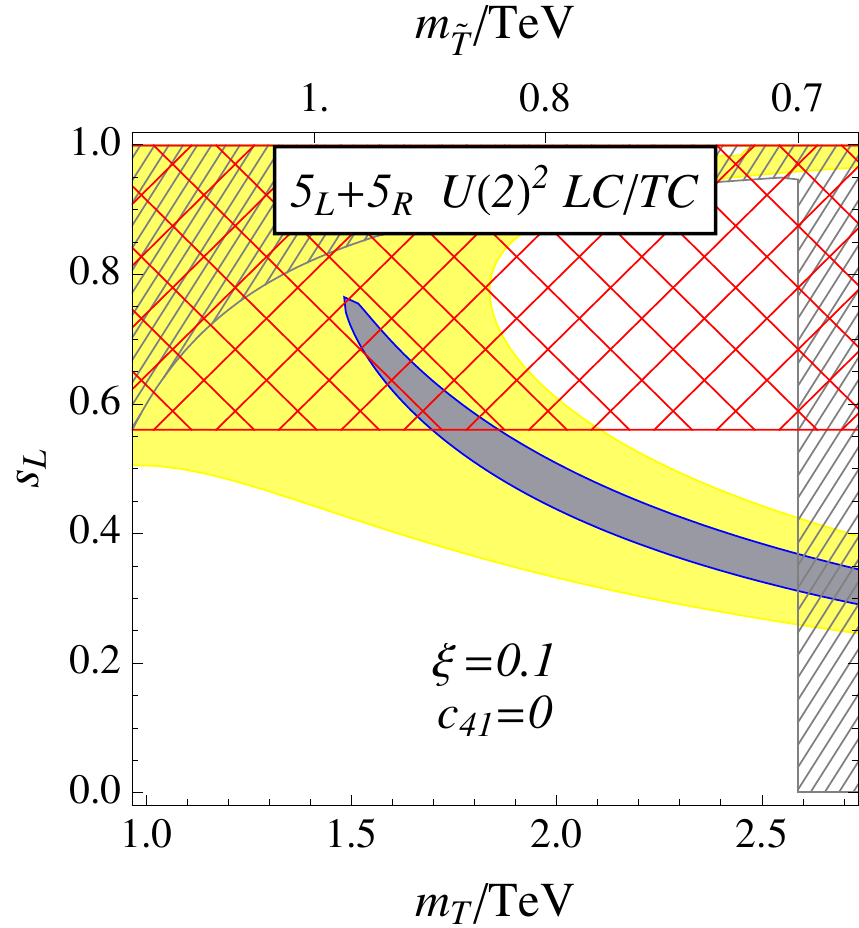}
\includegraphics[width=.32\textwidth]{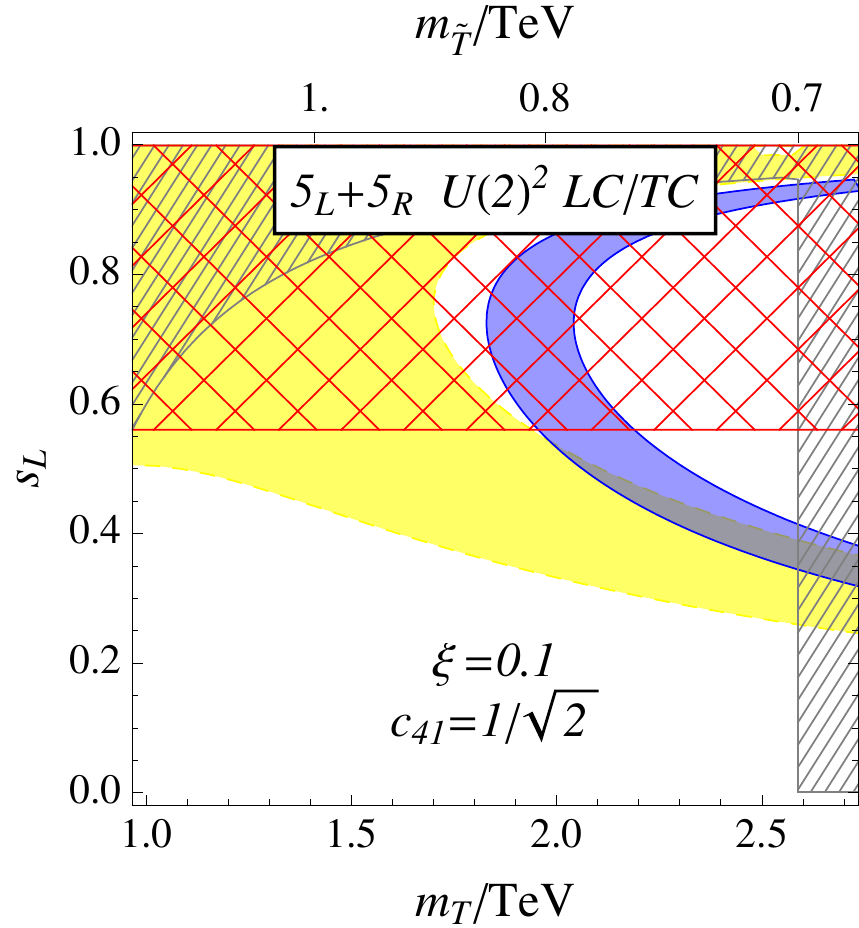}
\includegraphics[width=.32\textwidth]{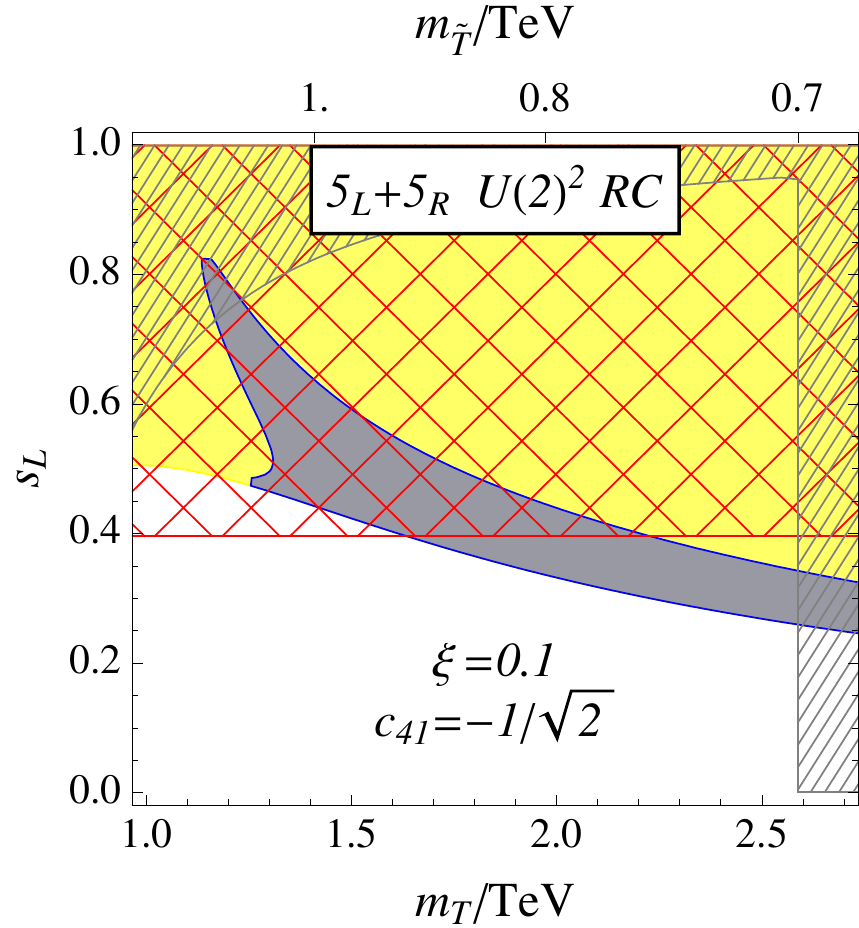}
\includegraphics[width=.32\textwidth]{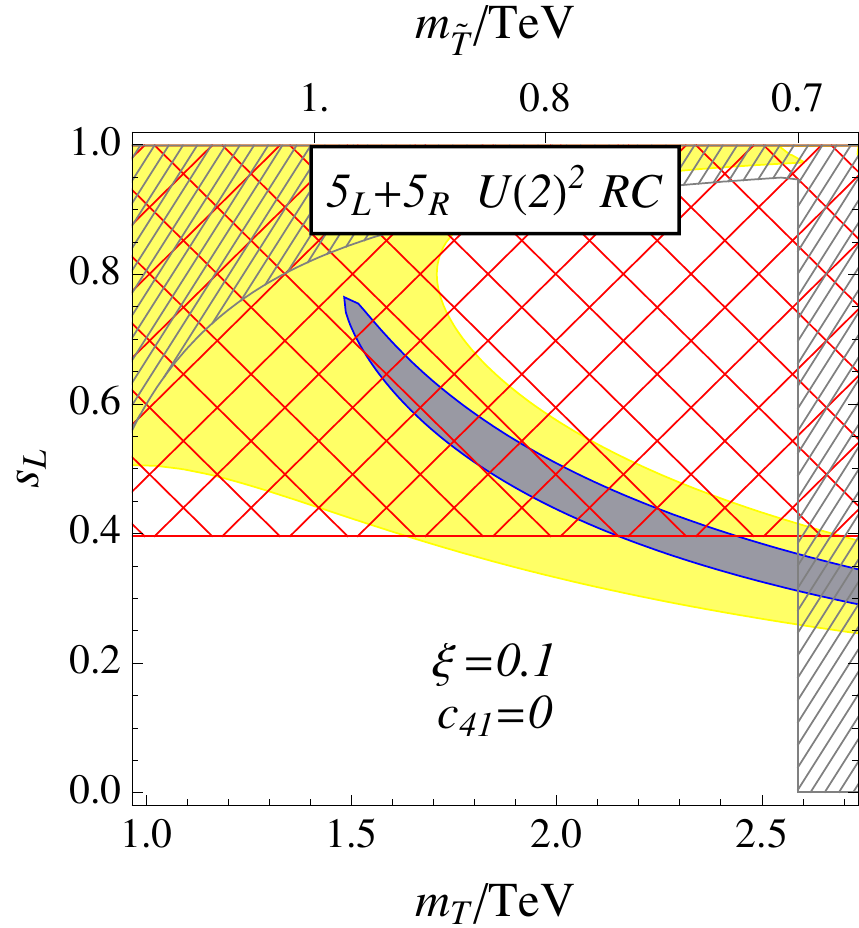}
\includegraphics[width=.32\textwidth]{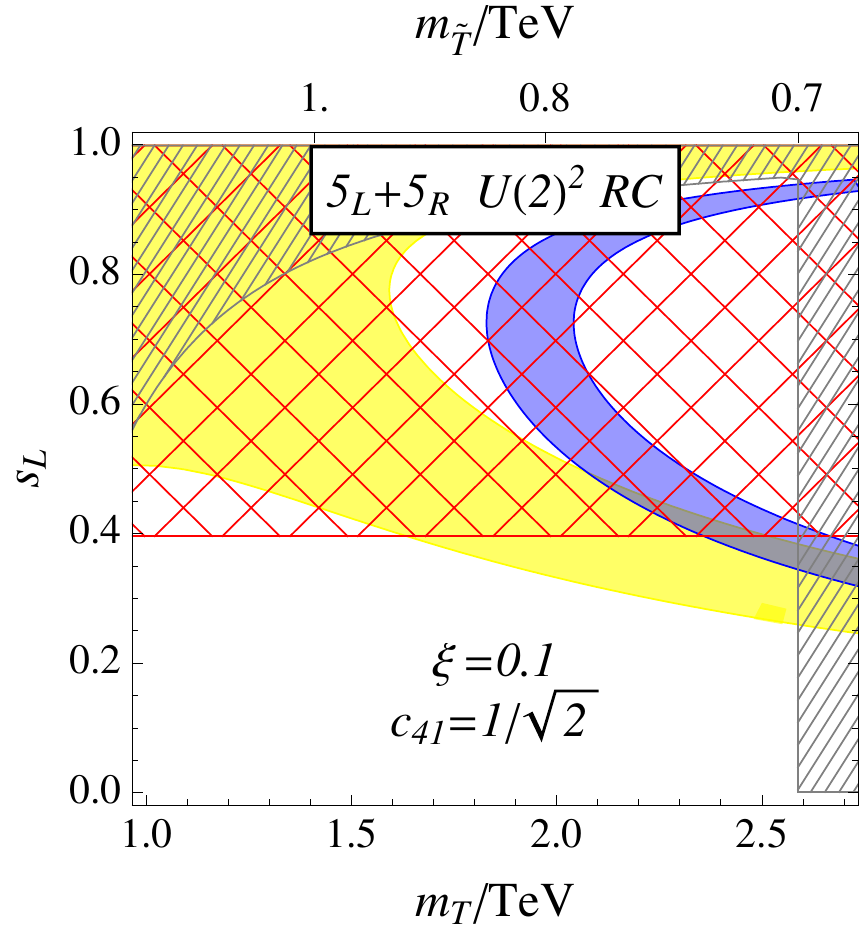}
\caption{Bounds on the \FLFR model with a fixed $m_h$ and $U(2)^2$ flavour symmetry, for $\xi=0.1$ and $c_{41}=-1/\sqrt 2, 0, 1/\sqrt 2$. Three upper plots correspond to the Left Compositeness and mixed TC scenario (with $r_b=1/2$), lower plots correspond to the Right Compositeness. The axes correspond to the masses of the fourplet ($m_T$) and the singlet ($m_{\widetilde T}$) top partners and to the degree the $t_L$ compositeness $s_L$.
Grey dashed area is forbidden by direct searches, red meshed area is forbidden by $\Delta F=2$ FCNC constraints, blue area is allowed by $\hat S$ and $\hat T$, yellow -- by  $\delta g_{b_L}$.}
\label{fig:excl_U2_5_5}
\end{figure}

We present the combined constraints on the \FLFR model on Fig.~\ref{fig:excl_U2_5_5} for $\xi=0.1$ and $c_{41}=-1/\sqrt 2, 0, 1/\sqrt2$, for LC and TC cases (upper plots) and RC case (lower plots), with the Higgs mass fixed by the relation~(\ref{eq:mh_2s}) imposed on top partners. The dashed grey and red areas are excluded by the direct searches for top partners and FCNC respectively, blue area passes EWPT constraints, and yellow area is compatible with our condition on $\delta g_{b_L}$. For the $\Delta F=2$ constraints on LC and TC scenarios we took $r_b=1/2$.
Accounting for these constraints we conclude that the typical feature of RC scenario is the relatively light singlet partner with $m_{\widetilde T} \lesssim 0.9$~TeV, as a consequence of a combination of the EWPT and $\Delta F=2$ bounds, with the maximal allowed mass achieved for negative $c_{41}$. We also performed a test with a slightly relaxed $m_h(m_T,m_{\widetilde T})$ in order to mimic the 3-site model, where both singlet and fourplet can be slightly heavier compared to what is predicted by Eq.~(\ref{eq:mh_2s}). This allowed to uplift the maximal $\widetilde T$, but the combination of EWPT and FCNC constraints still puts a limit $m_{\widetilde T}\lesssim 1.5$ TeV. 
One should mention that Eq.~(\ref{eq:mh_2s}) together with the requirement for the correct top mass typically allows for two solutions. The second one, which we prefer not to use in our analysis, requires $y_R f_{\pi} > m_1$, in which one can see a contradiction to the assumption about the small breaking of the Goldstone symmetry by the external perturbation. 
Taking this second solution we obtain the $m_{\widetilde T}$ not bounded from above.

The masses of the partners in LC and TC scenarios are practically unconstrained for our $r_b$ choice, the only interesting bound is $s_L \lesssim 0.6$. 
 

\subsection*{\FLOR}

\begin{figure}
\centering
\includegraphics[width=.49\textwidth]{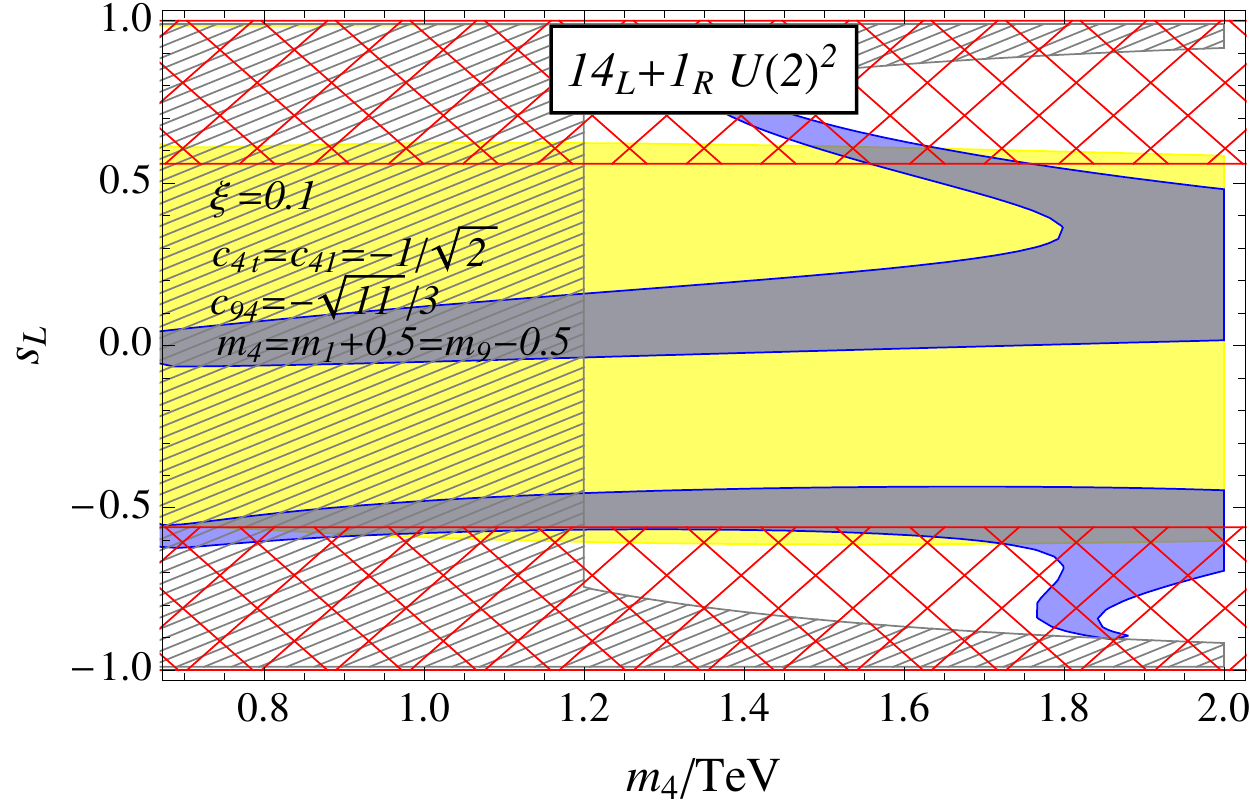}
\includegraphics[width=.49\textwidth]{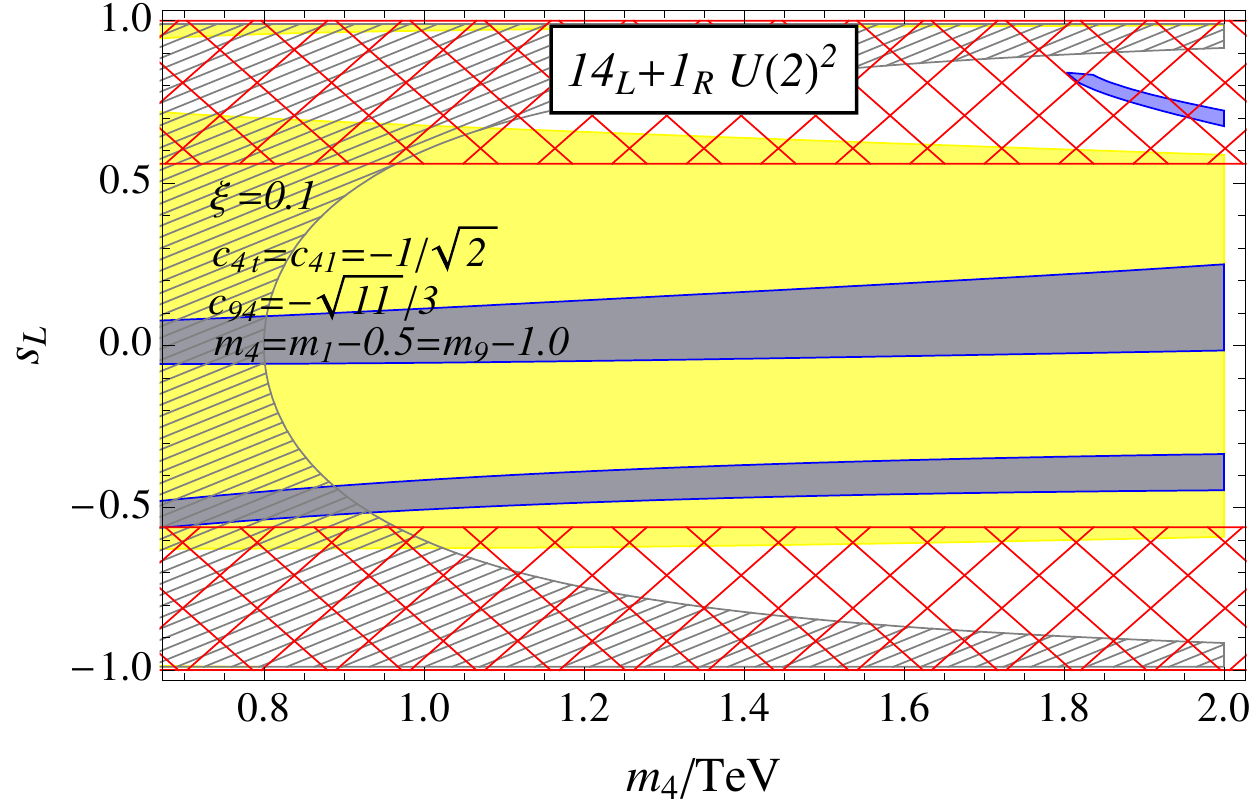}
\caption{Bounds on the \FLOR model with $U(2)^2$ flavour symmetry, for $\xi=0.1$, $c_{41}=c_{4t}=-1/\sqrt 2$, $c_{94}=-\sqrt{11}/3$, axes correspond to the fourplet mass parameter $m_4$ of the top sector and to the degree the $t_L$ compositeness $s_L$. All the mixings are set to be equal $y_{iL}=y_L = s_L m_T / f_{\pi}$. Grey dashed area is forbidden by direct searches, red meshed area is forbidden by $\Delta F=2$ FCNC constraints, blue area is allowed by $\hat S$ and $\hat T$, yellow -- by  $\delta g_{b_L}$. The mass parameters are chosen such that $m_4=m_1+0.5=m_9-0.5$ (left panel) or $m_4=m_1-0.5=m_9-1.0$ (right panel).}
\label{fig:excl_U2_14_1}
\end{figure}

The main difference of the \FLOR model with respect to the previously considered \FLFR is that the top mass is not generated via mixings with the massive composite resonances. Hence these mixings can be made small without conflicting with $m_{top}$. 
Another important difference is that the configuration with only the light fourplet is now allowed by the $\hat T$ parameter, provided that $c_{4t}$ and $s_{L4}$ are of the opposite signs~\cite{ewpt}. The new multiplet with respect to \FLFR, the ninplet, being light, can generate shifts in $\hat T$ and $g_{b_L}$ going to the right direction, which however are relatively small. Though the nineplet can hardly be the only light multiplet given that it induces a large positive shift in $\hat S$.   
To demonstrate the compatibility of the \FLOR model with the bounds we present two plots on Fig.~\ref{fig:excl_U2_14_1}, for  $m_1 < m_4 < m_9$ (left panel) and $m_4<m_1<m_9$ (right panel) with $0.5$~TeV mass gap between multiplets. We set equal all the mixings $y_{iL}=y_L=s_L m_T/f_{\pi}$, and the values of $d$-symbol coefficients are chosen to cancel the divergence in $\hat S$, $c_{41}=c_{4t}=-1/\sqrt 2$ and $c_{94}=-\sqrt{11}/3$. From this plots we see that the masses of the composite partners are constrained only by the direct searches, and $s_L$ in the chosen configuration can be as large as $0.6$, given the same $r_b$ choice as in the \FLFR case, $r_b=1/2$.

\section{Summary}
\label{sec_sum}

In this paper we have presented the analysis of the constraints imposed on Composite Higgs models by Flavour Physics observables in combination with ElectroWeak Precision Tests, direct searches for composite resonances and searches for the quark compositeness. 
We focused on the naturalness-related sector of the model, which generically induces the largest deviations of the observables with respect to the Standard Model predictions, and at the same time is expected to lie not too far above the electroweak scale. Hence our basic set-up included only the lightest levels of colored composite fermionic resonances which are  tightly related with the Higgs potential generation. To make the connection with the naturalness more precise we employed the CCWZ formalism, allowing to capture the most general features of the PNGB Higgs. One of the main organizing principles of our models was the Partial Compositeness paradigm in the top quark sector. For the up and charm sectors, besides the Partial Compositeness, we also considered a possibility to generate the flavour structures by means of Technicolor-like interactions.

We considered several possible ways to generate the SM flavour structures, all of them were based on an assumption of a presence of certain flavour symmetries in the strong sector. We have shown that in case of $U(3)$-symmetric models, the Goldstone symmetry implies a presence of certain regions in the parameter space where the strongest bounds on the scenarios can be relaxed. In case of the Left Compositeness this is the region where the singlet and the fourplet partners have similar mass, i.e. the region which will completely covered by direct searches in the last turn. The important test of this scenario thus can be the updated data from the searches for light quark compositeness. 
The Right Compositeness prefers the regions with the light singlet and heavy fourplet partners, thus the direct searches for composite singlets will cover the area allowed by the minimal CH models of this type quite soon. Additional tests of $U(3)$ scenarios will come with improved results of Higgs properties at the LHC, as the latter can be affected by the light quark compositeness~\cite{Delaunay:2013iia}. 
The viability of both LC and RC configurations however relies on the assumption of the presence of additional correlated contributions to the $\hat S$ or $\hat T$ parameters in case of LC, or to the Higgs potential in RC case. Both types of contributions can be generated by other sectors of the theory, not included in our analysis. As was discussed in Section~\ref{sec:S_param}, the $\hat S$ parameter can be strongly sensitive even to the physics which is not directly related to electroweak symmetry breaking. In Ref.s~\cite{Barnard:2013hka,Carmona:2014iwa} it was also shown that the Higgs mass can in certain cases allow for the heavier top partners than expected in the minimal set-ups.

\begin{figure}
\centering
\includegraphics[width=.49\textwidth]{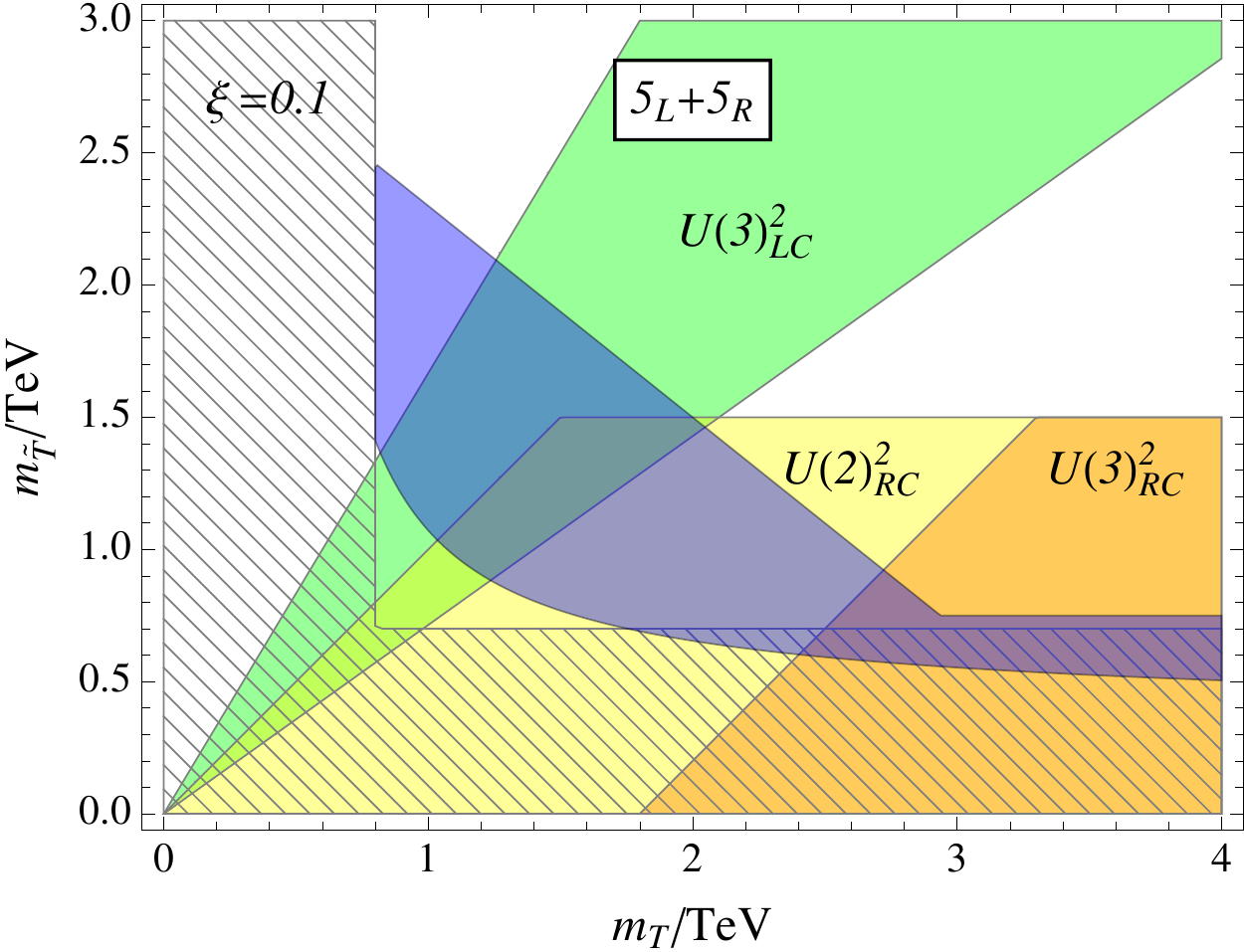}
\caption{Schematic summary plot showing the allowed space of the \FLFR model in the plane of top partner masses $m_T-m_{\widetilde T}$, preferred by $U(3)^2$ LC (green), $U(3)^2$ RC (orange) and $U(2)^2$ RC (yellow). The other considered flavour patterns can be allowed for any combination of $m_T$ and $m_{\widetilde T}$. The blue area approximately corresponds to the region of the 3-site model parameter space where one can reproduce the correct Higgs mass, assuming that it is dominated by the top sector and taking $m_T$ and $m_{\widetilde T}$ as the masses of the lightest $t_L$ and $t_R$ partners. Grey dashed area is excluded by direct searches.}
\label{fig:sum_5_5}
\end{figure}

The $U(2)$-symmetric scenarios expectedly have less constrained parameter space. Moreover, they allow for more freedom in choosing the mechanisms for the generation of the SM quark masses. We considered the models where the right-handed top quark is partially or totally composite state, while the light quarks are either partially composite or elementary with Technicolor-like Yukawa interactions. The most constrained are the models with partially composite light fermions and the RC flavour breaking scheme. They typically require a presence of light singlet partners with the mass less than $\sim1.5$~TeV. The masses of composite resonances in the rest of the scenarios  are dominantly constrained by the direct searches. This last conclusion however is not universal and depends on the details of the down-type quark sector, whose parameters were chosen to minimize the FCNC constraints. The least constrained $U(2)$-symmetric scenario is the one with the totally composite $t_R$, in which the top mass is not related to the mixings of $t_L$ with other composite multiplets.
On Fig.~\ref{fig:sum_5_5} we present the areas preferred by different flavour realizations in the \FLFR  scenario, in the plane of masses of the top partners.

Despite the fact that we preferred not to use the constraints imposed by $Z\bar bb$ couplings due to a large ambiguity and UV-dependence, we have identified the effect that can potentially allow for sizable deviations of the $Z$ couplings measured in $Z\to \bar bb$ decays, without affecting the measurements at lower energies, such as meson properties. This can be important since, for instance, for certain flavour patterns the new data on $BR(B_s \to \mu \mu)$  prefers no deviations from the SM predictions in $Z \bar b_Lb_L$ coupling, while the $Z\to \bar bb$ decays measurements can point towards certain distortions with respect to the Standard Model.

\subsubsection*{Acknowledgments}

The author is grateful to A.Tesi for collaboration on the early stages of this work, to R.Barbieri, D.Derkach, S.Rychkov and E.Vigiani for useful discussions and especially to A.Wulzer for many suggestions and comments. This work was supported by MIUR under the contract 2010 YJ2NYW-010 and in part by the MIUR-FIRB grant RBFR12H1MW.  




\end{document}